\documentclass[nohyper,notoc]{JHEP} 

\usepackage{epsfig}

\newcommand\fverb{\setbox\pippobox=\hbox\bgroup\verb}
\newcommand\fverbdo{\egroup\medskip\noindent%
                        \fbox{\unhbox\pippobox}\ }
\newcommand\fverbit{\egroup\item[\fbox{\unhbox\pippobox}]}
\newbox\pippobox

\title{
Lattice instanton action from 3D SU(2) Georgi-Glashow model
}

\author{
Tateaki Yazawa and   Tsuneo Suzuki
\\
Institute for Theoretical Physics, Kanazawa University, \\
Kanazawa 920-1192, Japan.\\
E-mail: \email{yazawa@hep.s.kanazawa-u.ac.jp}\\
E-mail: \email{suzuki@hep.s.kanazawa-u.ac.jp}}

\received{\today}               
\accepted{\today}               

\preprint{KANAZAWA ~2000-15}      

\abstract{
3D Georgi-Glashow model is studied on the lattice in the London limit 
in an infrared but an intermediate region before the screening appears. 
Abelian and instanton dominances are observed after abelian projections 
in a unitary gauge and roughly in the maximally abelian gauge. Using an inverse
 Monte-Carlo method, we determine an effective instanton action in both
 gauges. When  we restrict ourselves to some regions of parameters $\beta$
  and $\kappa$, 
we obtain an almost perfect instanton action,
performing a block-spin transformation on the dual lattice.
It takes a form of a  Coulomb gas and 
reproduces fairly well the string tension obtained analytically by
 Polyakov.  The almost perfect actions in both gauges look the same in
 the infrared region, which suggests gauge independence. 
}

\keywords{Solitons Monopoles and Instantons, Confinement,
Lattice Gauge Field Theories
}

\begin{document}

\section{Introduction}

It is very important to understand confinement mechanism of QCD.
Wilson's lattice formulation \cite{wilson} shows that the confinement
is a property of a non-Abelian gauge theory of strong interaction and
there are a lot of numerical lattice calculations 
showing the confinement of color.
But  the mechanism of confinement is still not well understood.

One of approaches to the confinement problem is to search
for relevant dynamical variables 
and to construct an effective theory in terms of these variables.
In 1970's,  it was pointed out 
that the confinement of quarks may be explained as the dual Meissner effect
due to condensation of monopoles
\cite{Mandelstam,tHooft-Meissner}.

In (QCD)$_4$, \mbox{'t Hooft}\cite{tHooft-Abelian-Prj} proposed an idea of 
abelian projection. After a partial gauge fixing (called as abelian projection),
SU(N) gauge theory can be reduced to an abelian U(1)$^{N-1}$ theory with $N-1$
different charges and magnetic charges. Namely, $N-1$ monopoles appear after 
the abelian projection.
Quark confinement could be understood by the dual Meissner effect
due to condensation of these monopoles.
Actually
there have been many numerical data supporting the above conjecture
when one performs the abelian projection in the Maximally abelian (MA) gauge.

The first interesting findings are some phenomena called abelian dominance.
Abelian Wilson loops composed of abelian link variables alone
seem to reproduce essential features of quark confinement in the
Maximally Abelian gauge
\cite{Suzuki-Yotsu}.

The second is  monopole dominance.
It has been shown that monopoles alone are responsible for 
abelian quantities in the infrared region.
Abelian Wilson loops and abelian Polyakov loops 
are written as a product of monopole and photon contributions.
The monopoles which are defined by 
DeGrand and Toussaint\cite{DeGrand-Toussaint}
reproduce the confinement features
\cite{Shiba-suzuki1}.

Furthermore  an effective
lattice monopole action was derived by Shiba and Suzuki\cite{Shiba-suzuki2} 
in a simple case with two-point interactions alone. The work was extended 
to include 4 and 6 point interactions in \cite{KKNS} and \cite{FKS}.

There is another system having such a magnetic quantity naturally.
It is the three-dimensional Georgi-Glashow model 
(GG)$_3$
where
there exist a
famous \mbox{'t Hooft}-Polyakov instanton 
\cite{tHooft-Polyakov-mn1,tHooft-Polyakov-mn2}
having a magnetic charge.
Polyakov showed analytically \cite{Polyakov77} that
the string tension of 3D Georgi-Glashow model has a finite value.
He made a quasi-classical calculation  
using a dilute Coulomb gas approximation of 
\mbox{'t Hooft}-Polyakov instantons.
However justification of his method is not clear.
If the quasi-classical assumption using the dilute 
instanton gas is true, abelian and instanton dominances should hold 
in the infrared region of three-dimensional Georgi-Glashow model.
It is expected naturally that an effective abelian instanton action 
works well in the infrared region.

It is shown in \cite{ambjorn} that Polyakov's picture of the dilute Coulomb 
instanton gas is too simple to describe all infrared physics of (GG)$_3$.
Actually, it can not explain the screening of doubly charged particle due to 
$W^{\pm}$ bosons. However, such a screening appears only for particles with 
even charge and also in the long infrared range of $O(2M_W/\sigma)$, where 
$M_W$ is a mass of the W boson and $\sigma$ is the string tension.
Hence if we restricted ourselves to an infrared but an intermediate region 
before the screening appears, we may be allowed to adopt such a picture 
as Polyakov's one.

It is just the purpose of this study 
to test the Polyakov's assumptions and to check calculations done by him 
quantitatively formulating the theory on the lattice 
and restricting ourselves to 
the intermediate region before the screening appears.
Our method is  Monte-Carlo numerical simulations without ad-hoc assumptions.
In Section {\bf 2},
we briefly review the Polyakov's work in 1977\cite{Polyakov77}.
We calculate the total energy of \mbox{'t Hooft}-Polyakov instanton
in the London limit.
In Section {\bf 3},
we show the method of our numerical study. 
We derive a lattice instanton action in 3D Georgi-Glashow model using an inverse
Monte-Carlo method. We then perform
a block-spin transformation of DeGrand-Toussaint instantons.
These discussions are almost similar to those in 4D pure QCD  in 
MA gauge fixing\cite{Shiba-suzuki2}.
In Section {\bf 4},
we show our numerical results of the lattice monopole action in 3D 
Georgi-Glashow model.
Then we calculate the string tension derived by Polyakov using the 
parameters determined in our numerical simulations.
Section {\bf 5} is devoted to concluding remarks.

\section{Confinement mechanism in 3D Georgi-Glashow model}
\label{GG}
We review briefly the \mbox{'t Hooft}-Polyakov instanton 
 and confinement mechanism due 
to the instantons
 as discussed by Polyakov in Ref\cite{Polyakov77}.

\subsection{Georgi-Glashow model and \mbox{'t Hooft}-Polyakov instanton}

The three-dimensional Georgi-Glashow model is given by
\begin{equation}
S=\int {\rm d}^3 x \left\{ \frac14 F_{\mu\nu}^a F_{\mu\nu}^a
  + \frac12 {\cal D}_\mu \phi^a {\cal D}_\mu \phi^a
  + \frac{\lambda}{4} (\phi^a \phi^a -v^2)^2 \right\},
\label{GGAction1}
\end{equation}
where
\begin{equation}
F^a_{\mu\nu}=\partial_\mu A^a_\nu - \partial_\nu A^a_\mu
  +g \epsilon^{abc}A^b_\mu A^c_\nu,
\label{fmn1}
\end{equation}
\begin{equation}
{\cal D}_\mu \phi^a = \partial_\mu \phi^a + g \epsilon^{abc}A^b_\mu\phi^c.
\label{Dphi1}
\end{equation}

It is a SU(2) Yang Mills theory with an adjoint Higgs field.
It has a phase with
spontaneously broken symmetry (SU(2) $\rightarrow$ U(1))
and the gauge field becomes massive because of the Higgs mechanism.
The gauge boson mass is given at the tree level by 
\begin{equation}
m_W \simeq gv.
\label{mW}
\end{equation}

In the case of three space-time dimensions,
the theory has an instanton solution
------ \mbox{'t Hooft}-Polyakov instanton.
One instanton solution takes the form \cite{tHooft-Polyakov-mn1}
\begin{eqnarray}
\phi^a &=& \frac{r^a}{gr^2}H(gvr),
\label{phi-H} \\
A^a_\mu &=& \frac{1}{g}\varepsilon_{a\mu b}\frac{r^b}{r^2}[1-K(gvr)],
\label{A-K}
\end{eqnarray}
where $H$ and $K$ are dimensionless functions determined by
the equations of motion.  They satisfy 
\begin{equation}
\left.
 \begin{array}{l}
  K(gvr) \longrightarrow 0  \\
  H(gvr) \longrightarrow gvr
 \end{array}
 \right\}
       r \rightarrow \infty .
   \label{KH-inf}
\end{equation}

For the ansatz given in Eqs.(\ref{phi-H}) and (\ref{A-K}),
the energy of the one-instanton system is given by
\begin{eqnarray}
S=\frac{4\pi v}{g} \int_{-\infty}^{\infty}
  \frac{{\rm d}\xi}{\xi^2}
\left[
  \xi^2 (\frac{{\rm d}K}{{\rm d}\xi})^2
 +\frac12(\xi\frac{{\rm d}H}{{\rm d}\xi}-H)^2
 +\frac12(K^2-1)^2 \right. \nonumber \\
\left. +K^2H^2+\frac{\lambda}{4g^2}(H^2-\xi^2)^2
\right] ,
\label{tH-P-xi-action}
\end{eqnarray}
where $\xi=gvr$.
This can be calculated numerically
solving equations of motion for  $K$ and $H$.

The value of $S$ at this minimum is expressed as
\begin{equation}
S_{\rm cl}=\frac{m_W}{g^2}\epsilon(\lambda/g) .
\label{Scl}
\end{equation}
The function $\epsilon(x)$ is slowly varying.
For example, 
it is known \cite{tHooft-Polyakov-mn1} that 
$\epsilon(0)=4\pi$.
In our case, we calculate Eq.(\ref{tH-P-xi-action}) 
in the London limit ($\lambda\rightarrow\infty$)
because we adopt the London limit in lattice calculations done later.
We obtain 
\begin{equation}
\epsilon(\infty)=4\pi\times 1.67 .
\label{epsilon_infty}
\end{equation}

\subsection{Instanton dilute gas model and Confinement}

Polyakov showed that, when the Higgs mass is large enough,
the infrared structure of the model 
in its Higgs phase is dominated by a dilute Coulomb gas of
\mbox{'t Hooft}-Polyakov instantons  with long-range
interactions ($r\gg {m_W}^{-1}$) which disorder the system.

When we consider quantum fluctuations around the instanton background,
we need to consider zero mode solutions the number of which are 
equal to that of conservation laws violated by the classical solution
\cite{Polyakov77}.

The partition function in the one-instanton case is given in the tree approximation 
by
\footnote{
Polyakov evaluated the integral beyond the tree level (Ref.\cite{Polyakov77}).
The one-instanton partition function in the one-loop approximation is given by 
$ 
Z_1=\int{\rm d}{\mbox{\boldmath $R$}}N^{3/2}
{\rm Det}^{1/2}({\cal D}^{\rm cl}_\mu {\cal D}^{\rm cl}_\mu
    -g^2 [\Phi_{\rm cl} , [ \Phi_{\rm cl} , ~] ~] )
  \exp{(-\sum_{n\neq 0} \log{\frac{\Omega_n}{\Omega_0}})} m_W^3
  e^{-S_{\rm cl}}.
$\\
But we  focus only on the  tree-level result for simplicity.
}
%
\begin{equation}
Z_1 \simeq \int{\rm d}{\mbox{\boldmath $R$}} N^{3/2} m_W^3 e^{-S_{\rm cl}},
\label{1-instanton-partition-f}
\end{equation}
\\
where $N$ is the normalization factor 
\begin{eqnarray}
N &=& {\rm tr}\int{\rm d}^3x
 \{ (F^{\rm cl}_{\lambda\mu})^2 
 + ({\cal D}^{\rm cl}_\lambda \Phi^{\rm cl})^2 \} 
\nonumber \\
&=&\frac{4\pi v}{g} \int{\rm d}\xi \frac{1}{\xi^2}
    \left[2\xi^2 \left(\frac{{\rm d}K}{{\rm d}\xi}\right)^2
    +(K^2-1)^2 
    +K^2H^2 +\frac12\left(\xi\frac{{\rm d}H}{{\rm d}\xi}-H\right)^2 \right]
    \nonumber \\
 &=&\frac{m_W}{g^2}\alpha(\lambda/g^2)  .
\label{N}
\end{eqnarray}
It is evaluated in the London limit ($\lambda\rightarrow \infty$)
as 
\begin{equation}
\alpha(\infty)\simeq 4\pi \times 3.33  .
\label{alpha_infty}
\end{equation}

Let us next discuss a system of several instantons.
In Ref.\cite{Polyakov77},
Polyakov considered a case in which
the positions of instantons are far from each other ( $r \gg m_W^{-1}$)
and the instantons have only the Coulomb interaction. This is justified
when the Higgs mass is large.
Then the action is given by
\begin{equation}
S=\frac{m_W}{g^2}\epsilon(\lambda/g^2)\sum_a q_a^2
 +\frac{2\pi}{g^2}\sum_{a\neq b}\frac{q_a q_b}{|\vec{x}_a-\vec{x}_b|}.
\label{polyakov-model-action}
\end{equation}
where $q_a$ is an instanton charge and $q_a =\pm 1 $ alone are taken into
account.

The partition function of the dilute gas model is given by
\begin{equation}
Z=\sum_{n,\{q_a\}}\frac{\zeta^n}{n!} \int \prod_{j=1}^n
  {\rm d}^3x_j \exp{ \left\{ -\frac{2\pi}{g^2}
    \sum_{a\neq b}\frac{q_a q_b}{|\vec{x}_a - \vec{x}_b|} \right\} } ,
\label{dilute-instanton-partition-func}
\end{equation}
where
\begin{equation}
\zeta=N^{3/2}{m_W}^3 
  e^{-\frac{m_W}{g^2}\epsilon(\lambda/g^2) } .
\label{zeta}
\end{equation}

This  is rewritten as 
\begin{eqnarray}
Z&=&
  \int {\cal D}\chi \sum_{n,\{q_a\}}\frac{\zeta^n}{n!}
 \int(\prod_{j=1}^n {\rm d}^3x_j)
 \exp{ \left\{
   -\frac12\left(\frac{g}{4\pi}\right)^2\int{\rm d}^3x(\partial_\mu \chi)^2
     +i\sum_a q_a \chi(x_a) \right\} } \nonumber \\
 &=&\int{\cal D}\chi \exp{\left\{-\frac12\left(\frac{g}{4\pi}\right)^2
 \int{\rm d}^3x((\partial_\mu \chi)^2 - 2M^2 \cos{\chi} ) \right\} } ,
\end{eqnarray}
where
\begin{equation}
M^2=2\left(\frac{4\pi}{g}\right)^2 \zeta .
\label{M}
\end{equation}

The static potential between a pair of static electric charges is 
estimated by 
the Wilson loop (Fig.\ref{wilson-figure}) defined by
\begin{equation}
W[C]=\langle e^{ig\oint_C A^3_\mu {\rm d}x_\mu} \rangle
=\langle e^{ig \int_S B_\mu {\rm d}S_\mu } \rangle ,
\label{wilson_loop1}
\end{equation}
where the magnetic field from an instanton is
\begin{equation}
B_\mu(x)=\frac{1}{g} \frac{r_\mu}{r^3}.
\label{B-1mono}
\end{equation}
One easily recognizes in (\ref{B-1mono}) the Dirac quantization condition.
In the case of several instantons,
the strength of the magnetic field is given by
\begin{equation}
B_\mu(x)=\frac{1}{g} \int {\rm d}^3 y \frac{x_\mu-y_\mu}{|x-y|^3} k(y),
\label{B-mono}
\end{equation}
where the instanton density is
\begin{equation}
k(x)=\sum_a q_a \delta^3(x-x_a).
\label{mono-density}
\end{equation}
The Wilson loop  Eq.(\ref{wilson_loop1}) is expressed as 
\begin{equation}
W =\langle e^{i\int{\rm d}^3x \eta(x)k(x)} \rangle
    =\frac{Z[\eta]}{Z[0]} ,
\label{wilson_loop2}
\end{equation}
where
\begin{eqnarray}
&&Z[\eta]=\int D\chi\exp{\left\{ -\frac12\left(\frac{g}{4\pi}\right)^2 \int{\rm d}^3x
  [(\partial_\mu(\chi-\eta) )^2 - 2M^2 \cos{\chi}] \right\} }, 
\label{partition_eta} 
\\
&&\eta(x)=\int_S {\rm d}\vec{S}\cdot\frac{\vec{x}-\vec{y}}{|\vec{x}-\vec{y}|^3}
.
\label{eta}
\end{eqnarray}
\FIGURE{
\epsfig{file=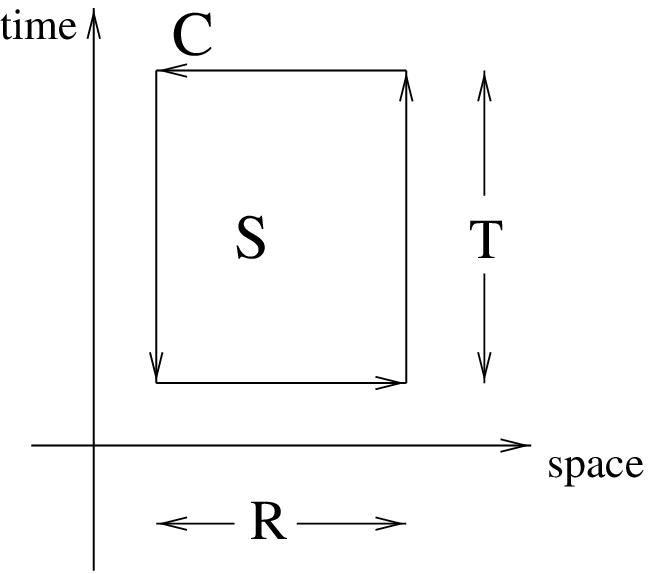,width=9cm,height=6cm}
\caption{Wilson loop}
\label{wilson-figure}
}

The path integral (\ref{partition_eta}) 
may be evaluated by the saddle-point solution
(see Appendix \ref{string_tension}):

\begin{eqnarray}
&&W[C] \sim \exp\left\{
-RT \times  \right. \\ \nonumber
  && \left. M\left(\frac{g}{\pi}\right)^2 \int_0^{\infty}{\rm d}z
  \left[\arctan\left(\frac{M^2RT}{2z\sqrt{(MR)^2+(MT)^2+4z^2}} \right)
    \frac{e^{-z}(1-e^{-2z})}{(1+e^{-2z})^2} \right] \right\}.
\label{wilson_st_math}
\end{eqnarray}
It shows an area law in the approximation of large $T$ and large $R$.
The string tension is obtained explicitly as
\begin{equation}
\sigma_{\rm cl}=\frac{M g^2}{4\pi}.
\label{st_an}
\end{equation}

Finally in this section, we show that 
 Eq.(\ref{polyakov-model-action}) is rewritten 
in the outside of 
the instantons as
\begin{equation}
S\sim\int {\rm d}^3x {\rm d}^3x' k(x)\frac{8\pi^2}{g^2}
\Delta^{-1}(x-x') k(x').
\label{polyakov-model-action2}
\end{equation}
We will compare later this form of action with a lattice instanton action
derived numerically.

\section{Lattice Georgi-Glashow model and instantons}
\label{method}

Now let us make an analysis of the Georgi-Glashow model 
in the framework of lattice field theory
without recourse to the quasi classical 
approximation. First we check the abelian dominance and the instanton
dominance in the infrared region numerically. 
Next we try to derive an effective instanton action in the continuum
limit and compare it with the action derived by Polyakov 
(\ref{polyakov-model-action2}).

\subsection{Lattice Georgi-Glashow model}

The lattice form of Eq.(\ref{GGAction1}) is expressed by
\begin{eqnarray}
S&=&\beta\sum_{x,\mu > \nu}
 \left(1-\frac12 {\rm tr}U_{\mu\nu}(x)\right) \nonumber \\
 & &\mbox{}
 +\kappa\sum_{x,\mu} 2
 {\rm tr}(\Phi_L(x)\Phi_L(x)-\Phi_L(x)U_\mu(x)\Phi_L(x+\mu)U_L^\dagger(x))
\nonumber \\ & & \mbox{}
 +\frac{\lambda_L}{4}\kappa^2\sum_x[2 {\rm tr}(\Phi_L(x)\Phi_L(x)) - 1]^2 ,
\end{eqnarray}
where the plaquette gauge variable is
\begin{equation}
U_{\mu\nu}(x)=U_\mu(x)U_\nu(x+\mu)U_\mu^\dagger(x+\nu)
U_\nu^\dagger(x) .
\end{equation}
In the case of three space-time dimensions, 
the relations between the continuum and the lattice quantities are 
\begin{eqnarray}
&& U_\mu(ax)=\exp[iag A^a_\mu(x)\frac{\sigma^a}{2}],
 ~~~~ \Phi_L(ax)=a\phi^a(x)\frac{\sigma^a}{2},
\\
&& m_L^2=a^2 v^2 \lambda , ~~~~ \lambda_L=a \lambda , \\
&& \beta=\frac{4}{g^2 a}, ~~~ \kappa^2=v^2 a .
\end{eqnarray}
For simplicity, we consider only the London limit 
$\lambda_L \rightarrow \infty $.
Then
\begin{equation}
2 {\rm tr}(\Phi_L(x)\Phi_L(x)) \longrightarrow 1 
\end{equation}
and
\begin{equation}
\sum_a\varphi_L^a\varphi_L^a = 1 .
\end{equation}
The action is reduced to
\begin{equation}
S=\beta\sum_{x,\mu > \nu}\left(1-\frac12 {\rm tr}U_{\mu\nu}(x)\right)
 +\kappa\sum_{x,\mu}\left(
  1-\frac12 {\rm tr}\{ \varphi_L^a(x)\sigma^a U_\mu(x) \varphi_L^b(x+\mu)
  \sigma^b U_\mu^\dagger(x) \} \right) .
\label{GGAction_lat-london}
\end{equation}

\subsection{The method}

The method of our numerical study is the following:
\begin{enumerate}
 \item 
We generate thermalized vacuum configurations  $\{U_\mu(x)\}$,
adopting,  for simplicity, a unitary gauge condition 
\begin{equation}
\varphi_L^a \equiv \delta^{a3} .
\label{unitary-g}
\end{equation}
In this gauge, the Higgs field disappears in the action. There remains
still $U(1)$ symmetry under the unitary gauge condition and hence 
it is regarded as one of abelian projections.
 \item 
The lattice considered is $48^3$ for $\beta=4.0\sim 6.0$ 
and $\kappa=1.00\sim 1.30$.  We also used $12^3$, $16^3$, $24^3$ 
and $32^3$ lattices to study  volume dependence.
 \item 
We next extract abelian link variables 
$\{u_\mu(x)\}$ in the unitary gauge as 
\begin{eqnarray}
U_\mu(x)&=&C_\mu(x)u_\mu(x)\\
&=&C_\mu(x)\left( \begin{array}{cc}
     e^{i\theta_\mu(x)} & 0 \\ 0 & e^{-i\theta_\mu(x)} 
                        \end{array} \right).
\label{ab-p}
\end{eqnarray}
 \item 
We generate also the vacuum configurations in the maximally abelian 
(MA) gauge from those in the unitary gauge. 
They are generated after  a gauge transformation in such a
way as  maximizing a quantity 
$
R=\sum_{x,\mu} {\rm tr}
\left( \sigma_3 U_\mu(x) \sigma_3 U^{\dagger}_\mu(x) \right).
$ 
The abelian link variables in MA gauge are extracted similarly as in 
(\ref{ab-p}) .
The Higgs field reappear in MA gauge, but we don't need them, since 
we evaluate here only operators composed of abelian link variables.
 \item 
An integer instanton charge is defined from the $U(1)$ plaquette
       variables following  DeGrand and Toussaint
       \cite{DeGrand-Toussaint}.
The U(1) plaquette variables are written by
\begin{equation}
\Theta_{\mu\nu}(x) \equiv 
 \theta_\mu(x)+\theta_\nu(x+\mu)-\theta_\mu(x+\nu)-\theta_\nu(x),
 (-4\pi<\Theta_{\mu\nu}(x)<4\pi).
\end{equation}
It is decomposed into two terms:
\begin{equation}
\Theta_{\mu\nu}(x) \equiv \bar{\Theta}_{\mu\nu}(x)+2\pi n_{\mu\nu}(x),
(-\pi\leq\bar{\Theta}_{\mu\nu}(x)<\pi).
\end{equation}
Here, $\bar{\Theta}_{\mu\nu}(x)$ is interpreted as the electro-magnetic flux
through the plaquette and the integer $n_{\mu\nu}(x)$ corresponds to the 
number of Dirac string penetrating a plaquette.
One can define a quantized instanton charge as 
\begin{eqnarray}
k(x)&=&\frac{1}{2\pi}\cdot\frac12 \epsilon_{\mu\nu\rho}\partial_\mu
    \bar{\Theta}_{\nu\rho}(x) \nonumber \\
  &=& -\frac12 \epsilon_{\mu\nu\rho} \partial_\mu n_{\nu\rho}(x) ,
\label{DG-instanton}
\end{eqnarray}
where $\partial$ denotes the forward difference on the lattice.
We then obtain vacuum ensembles $\{k(x)\}$  from the vacuum configuration
$\{u_{\mu}(x)\}$ both in unitary  and MA gauges.
 \item 
We determine an effective instanton action $S[k]$ from the 
vacuum ensemble $\{k(x)\}$ with the help of an inverse Monte-Carlo method first
developed by Swendsen\cite{Swendsen}
and extended to closed monopole currents by Shiba and Suzuki
\cite{Shiba-suzuki2}. Here instantons are a site variable and the 
original Swendsen method works. 
We consider a set of independent  instanton interactions
that are summed up over the whole lattice.
We denote each operator as $S_i[k]$.
Then the instanton action may be written as a linear combination of these
operators:
\begin{equation}
S[k]=\sum_i G_i S_i[k],
\label{instanton-action1}
\end{equation}
where $G_i$ are
coupling constants.

We determine the set of couplings $G_i$ from the instanton ensemble
$\{ k(x) \}$.
Practically, we have to restrict the number of interaction terms.
It is natural to assume that instantons that are far apart do not
interact strongly and to consider only short-ranged interactions of
instantons.
This assumption is not justified when the screening appears. Non-local 
long-range interactions are inevitable if we express the theory 
in terms of abelian quantities alone.
But here we are restricted to the intermediate region before 
the screening appears. 

The form of actions adopted here is 10 quadratic interactions :
\begin{eqnarray}
S[k] &=& G_0\sum_x k(x)^2 + G_1 \sum_{x,\mu} k(x)k(x+\mu) + G_2 \sum ...
 +G_9 \sum...
\\
 &=& \sum_{x,x'} k(x) D(x-x') k(x') .
\label{lat-instanton-action}
\end{eqnarray}
The detailed form of interactions is shown 
in Fig.\ref{action-operator}.

\FIGURE{
\epsfig{file=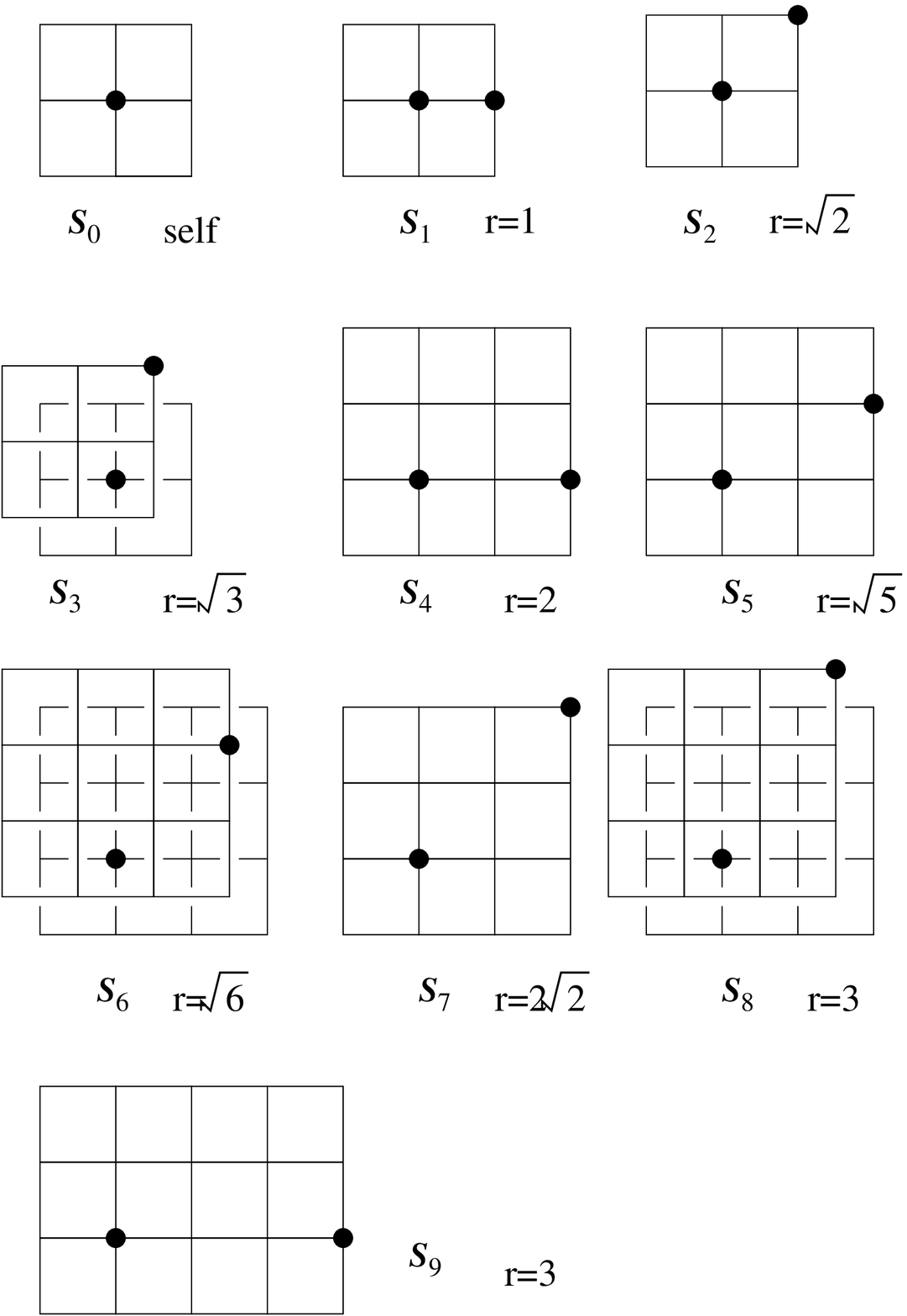,height=15cm}
\caption{Operators of instanton action}
\label{action-operator}
}
 \item 
To study the continuum limit, 
we perform a block-spin transformation in terms of the instantons.
We adopt $n=1,2,3,4,6,8$ extended instantons which are defined as
\begin{equation}
k^{(n)}(x^{(n)})=\sum_{i,j,k=0}^{n-1}k(nx^{(n)}+i\hat{1}+j\hat{2}+k\hat{3}) .
\label{extend-instanton}
\end{equation}
This is a block-spin transformation on the dual lattice and 
the renormalized lattice spacing is $b=na(\beta,\kappa)$.
The value of the blocked instanton  $k^{(n)}(x^{(n)})$ 
is integer.

We determine the effective instanton action from the blocked instanton
ensemble $\{ k^{(n)}(x^{(n)}) \}$ using the above inverse Monte-Carlo 
method.
Then one can obtain the renormalization flow in the coupling constant
space.

\end{enumerate}

\section{Results}

\subsection{String tension}

We first measure the static potential between a pair of quarks and the 
lattice string tension for $\beta=4.0\sim 6.0$ and $\kappa=1.00\sim 1.30$.
They are measured from  $T\rightarrow\infty$ of  $R\times T$  Wilson loops.
The Coulomb potential is not $1/R$ but $\log R$ 
because we are considering the three-dimensional case.
The static potential is fitted as 
 $A\log{R} + \sigma_L R + B$ (where $A, B$ are constants).
The usual $SU(2)$ Wilson loops and the string tension are evaluated to
fix the scale of this theory. 

Abelian Wilson loops in terms of the abelian link variables 
$\theta_\mu(x)$ in Eq.(\ref{ab-p}) are measured similarly.
The string tension is obtained from the abelian Wilson loops.
They are rewritten as 
\begin{equation}
W[C]=\exp\{i\sum_x \theta_\mu(x) J_\mu(x) \}
=\exp\{-\frac{i}{2}\Theta_{\mu\nu}(x)M_{\mu\nu}(x)\} ,
\end{equation}
where $J_\mu(x)=\partial'_\mu M_{\mu\nu}(x)$
.
Then the abelian Wilson loop operator can be decomposed into 
instanton and photon parts as shown in Ref.\cite{Shiba-suzuki1}:
Using an identity
\begin{equation}
M_{\mu\nu}(x)=-\Delta_L^{-1}(x-x')
 [\partial'_\alpha (\partial_\mu M_{\alpha\nu}-\partial_\nu M_{\alpha\mu})
 +\frac12 \epsilon_{\alpha\mu\nu} \epsilon_{\lambda\rho\sigma}
  \partial'_\alpha \partial_\lambda M_{\rho\sigma} ],
\end{equation}
we get
\begin{eqnarray}
W[C] &=& W_p[C] \cdot W_m[C], \\
W_p[C] &=& \exp\{-i\sum_{x,x'}\partial'_\mu \Theta_{\mu\nu}(x)
   \Delta_L^{-1}(x-x')J_\nu(x') \}, \\
W_m[C] &=& \exp\{ 2\pi i \sum_{x,x'}k(x)\Delta_L^{-1}(x-x')
  \frac12 \epsilon_{\alpha\rho\sigma}\partial_\alpha M_{\rho\sigma} \},
\end{eqnarray}
where $\Delta_L^{-1}(x)$ is the lattice Coulomb propagator.

In Table.\ref{st_L} and in Fig.\ref{q-q-pot}
we show the data of the string tensions for the regions of 
$\beta$ and $\kappa$.
It is interesting that the instanton contribution to the abelian Wilson
loop gives us a  static potential which is almost linear:
$V(R)\sim \sigma_L R + B$.  The string tension is determined 
also from this potential.
Except for the data of instanton contributions in MA gauge
which are smaller than
the $SU(2)$ ones, we see abelian and instanton 
dominances are fairy good. This supports the 
analysis of Polyakov\cite{Polyakov77}.
\TABLE{
\begin{tabular}{|r|r|r|r|r|r|r|} \hline
~$\beta$~ & ~$\kappa$~~~ & $\sigma_L$ {\small (SU(2))} 
  & $\sigma_L$ {\footnotesize (Abelian }& $\sigma_L$ {\footnotesize (Instanton }
  & $\sigma_L$ {\footnotesize (Abelian }& $\sigma_L$ {\footnotesize (Instanton }
\\
  &  &    
  & {\footnotesize ,Unitary G. )} & {\footnotesize , Unitary G. )}  
  & {\footnotesize , MA G. )}     & {\footnotesize , MA G. ) }
\\ \hline \hline

 6.0    & 1.030  &  0.0248(93)~ 
& 0.0213(74)~ & 0.0267(33) & 0.0238(39) & 0.0123(7)~ \\ \hline
 5.8    & 1.030  &  0.0239(91)~ 
& 0.0221(81)~ & 0.0265(29) & 0.0252(39) & 0.0129(4)~ \\ \hline
 5.6    & 1.035  &  0.0249(105) 
& 0.0236(95)~ & 0.0270(32) & 0.0285(43) & 0.0136(9)~ \\ \hline
 5.4    & 1.055  &  0.0291(109) 
& 0.0256(91)~ & 0.0290(32) & 0.0298(45) & 0.0155(3)~ \\ \hline
 5.2    & 1.065  &  0.0328(115) 
& 0.0234(106) & 0.0300(34) & 0.0288(49) & 0.0166(5)~ \\ \hline
 5.0    & 1.065  &  0.0288(129) 
& 0.0267(119) & 0.0304(32) & 0.0330(53) & 0.0185(6)~ \\ \hline
 4.8    & 1.090  &  0.0310(152) 
& 0.0318(108) & 0.0330(37) & 0.0402(61) & 0.0220(5)~ \\ \hline
 4.6    & 1.140  &  0.0405(173) 
& 0.0373(140) & 0.0393(41) & 0.0387(68) & 0.0256(10) \\ \hline
 4.4    & 1.170  &  0.0382(226) 
& 0.0425(182) & 0.0441(35) & 0.0449(78) & 0.0306(12) \\ \hline
 4.2    & 1.210  &  0.0429(287) 
& 0.0508(190) & 0.0509(39) & 0.0486(89) & 0.0361(13) \\ \hline
 4.0    & 1.255  &  0.0481(309) 
& 0.0596(221) & 0.0587(55) & 0.0554(99) & 0.0438(17) \\ \hline
\end{tabular}
\caption{Lattice String tension of Georgi-Glashow model}
\label{st_L}
}

\FIGURE{
\epsfig{file=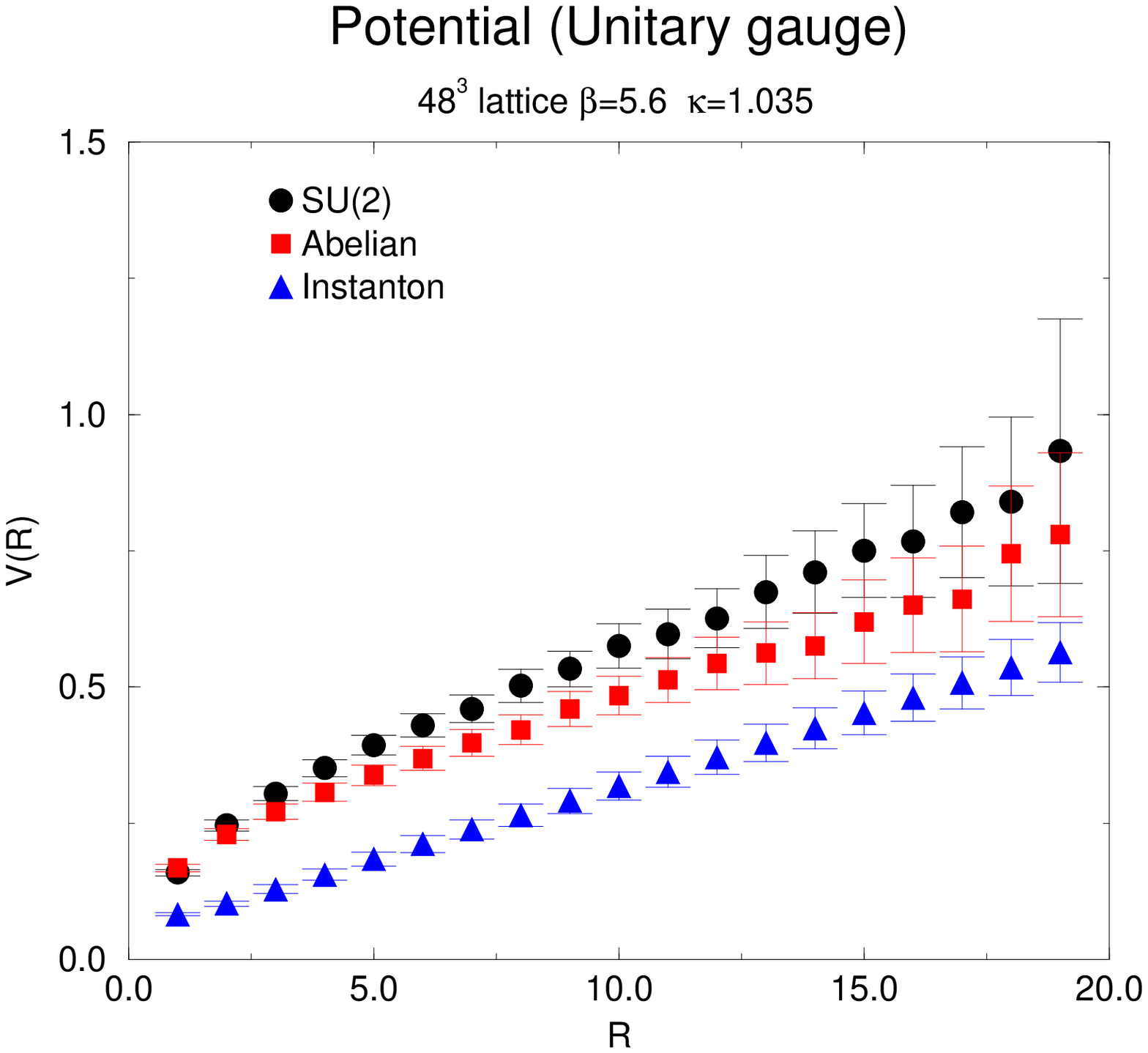,height=6cm}
\epsfig{file=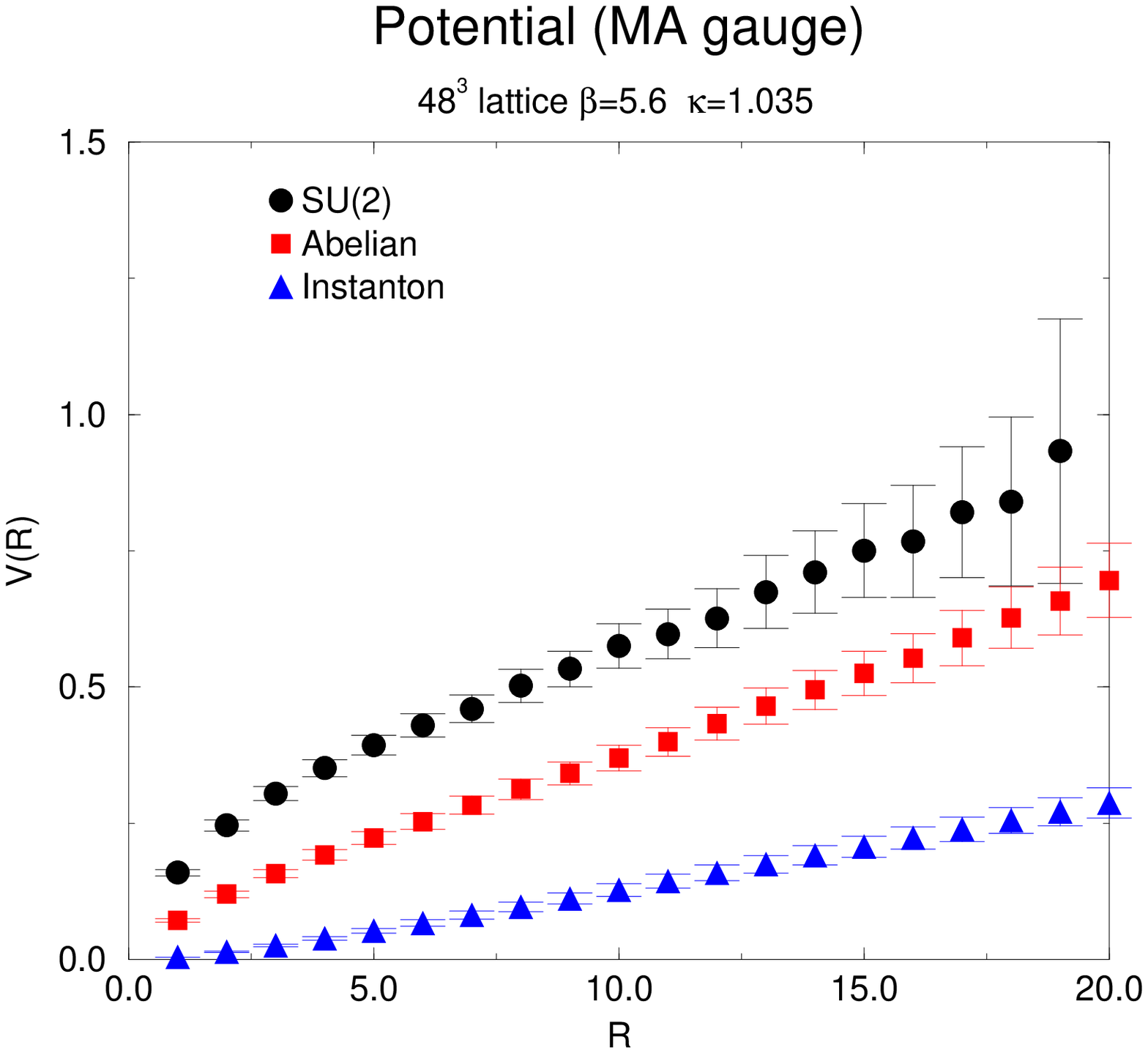,height=6cm}
\caption{$q$-$\bar{q}$ potential}
\label{q-q-pot}
}

\FIGURE{
\epsfig{file=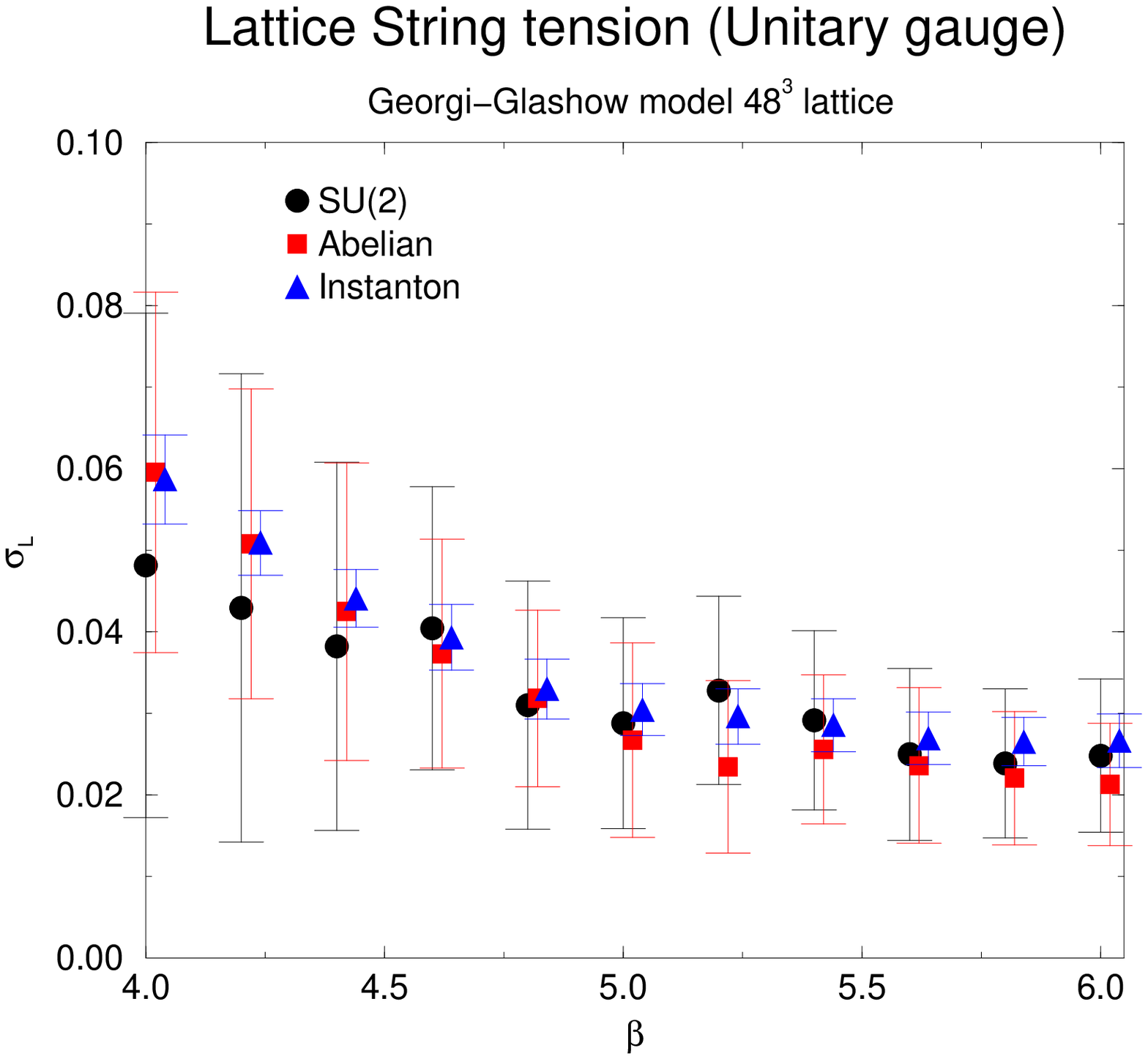,height=6cm}
\epsfig{file=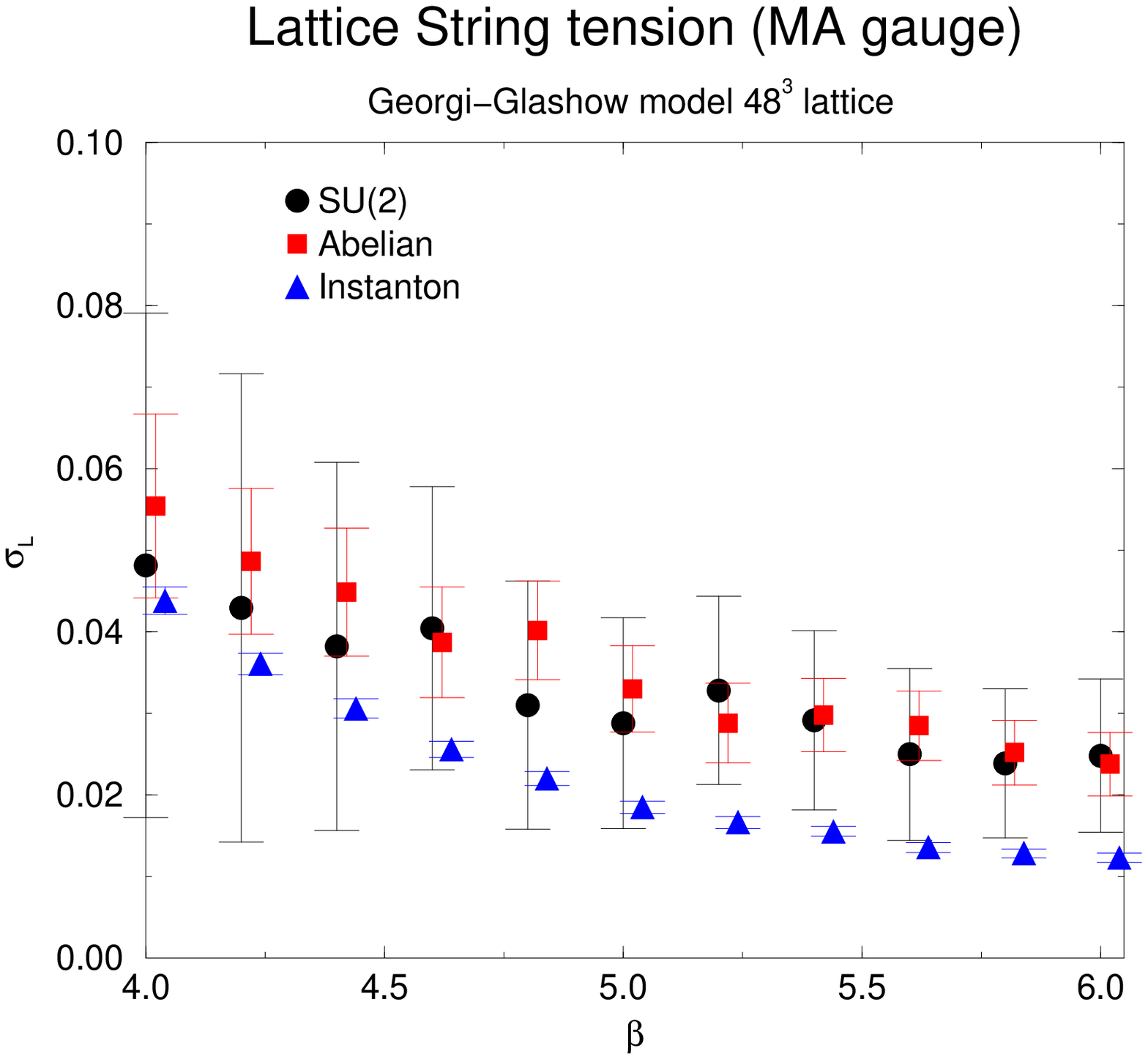,height=6cm}
\caption{Lattice String tension of Georgi-Glashow model}
\label{st_L-gr}
}

\subsection{Instanton vacuum in both gauges}

Polyakov\cite{Polyakov77} assumed the dilute instanton gas
approximation.  
Let us see  the instanton  vacuum configurations.
As we mention in the previous section,
 we calculate the instanton configuration using the DeGrand-Toussaint definition. 
The instanton configuration and anti-instanton configuration
 in unitary gauge condition for $\beta=4.0$ and $\kappa=1.30 $ 
are shown in Fig.\ref{mono-conf}.  
Those in MA gauge condition are shown in Fig.\ref{mono-conf-MA}.
These configurations look to be a dilute gas for this set of coupling constants.
\FIGURE{
\epsfig{file=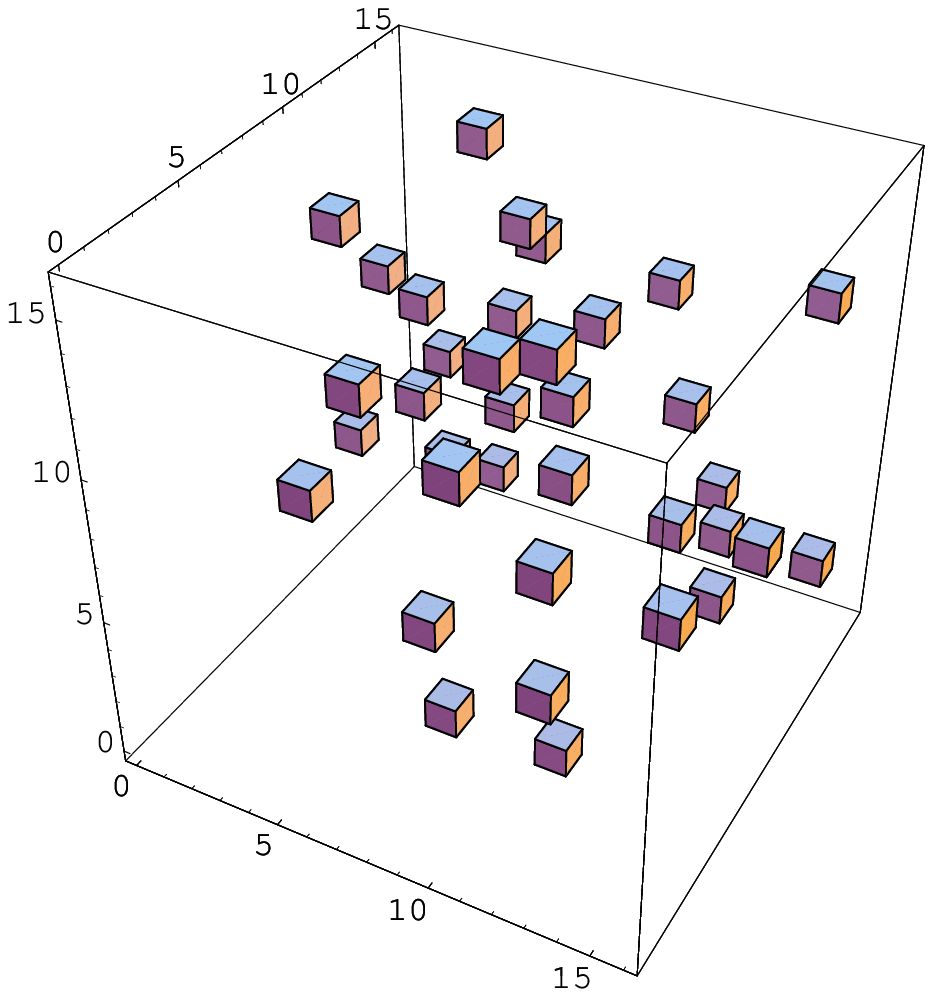,height=5cm}
\epsfig{file=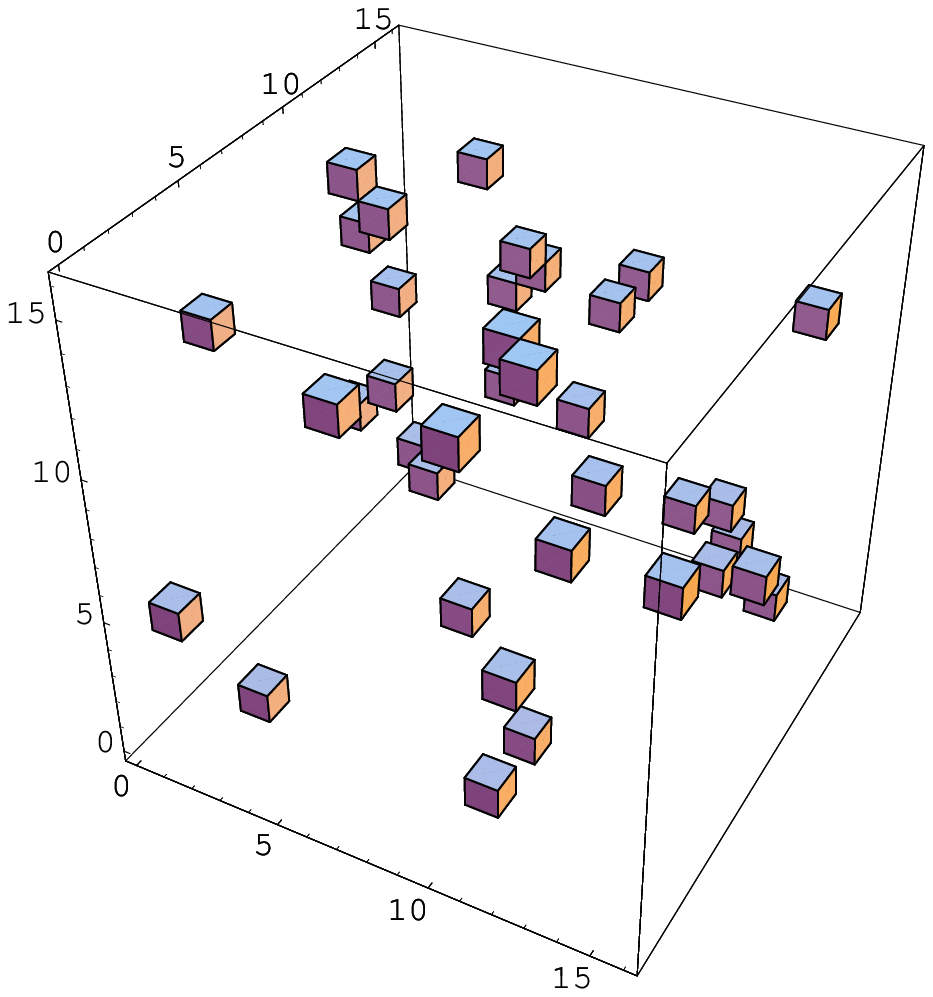,height=5cm}
\caption{Instanton (left) and anti-instanton (right) 
vacuum configuration (unitary gauge)}
\label{mono-conf}
}
\FIGURE{
\epsfig{file=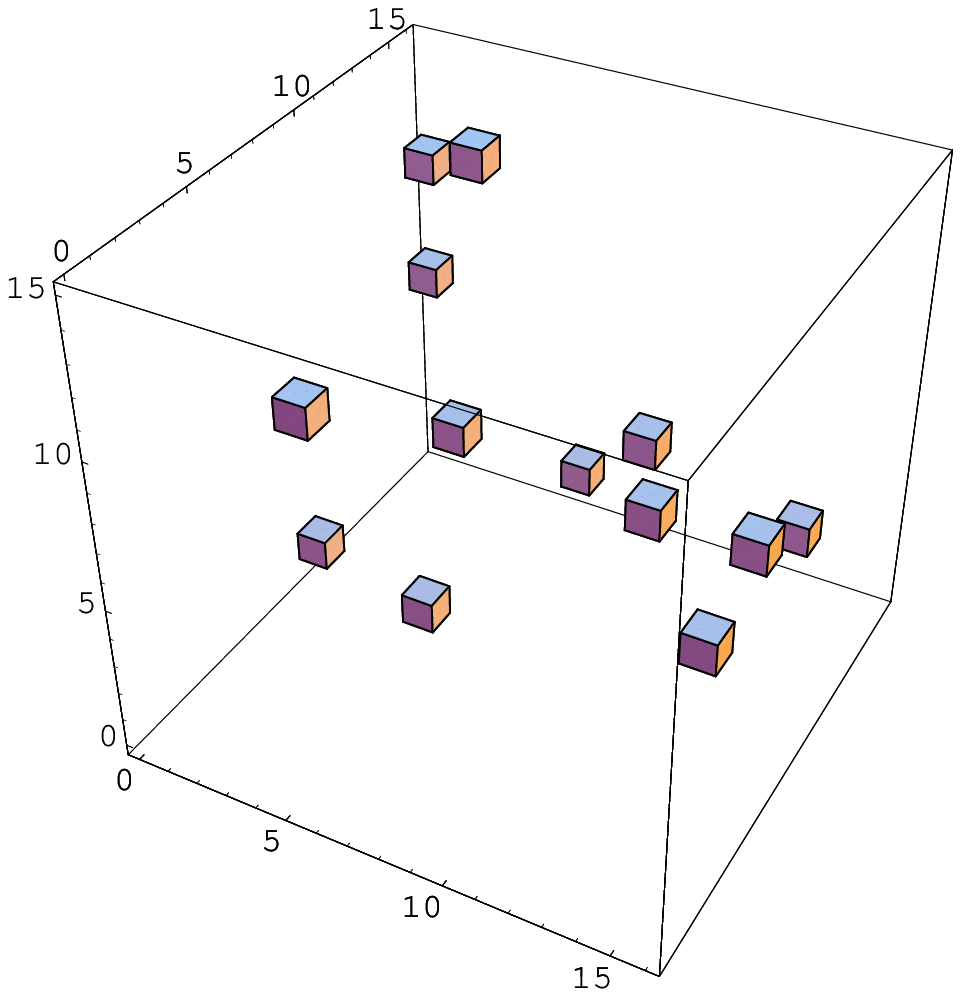,height=5cm}
\epsfig{file=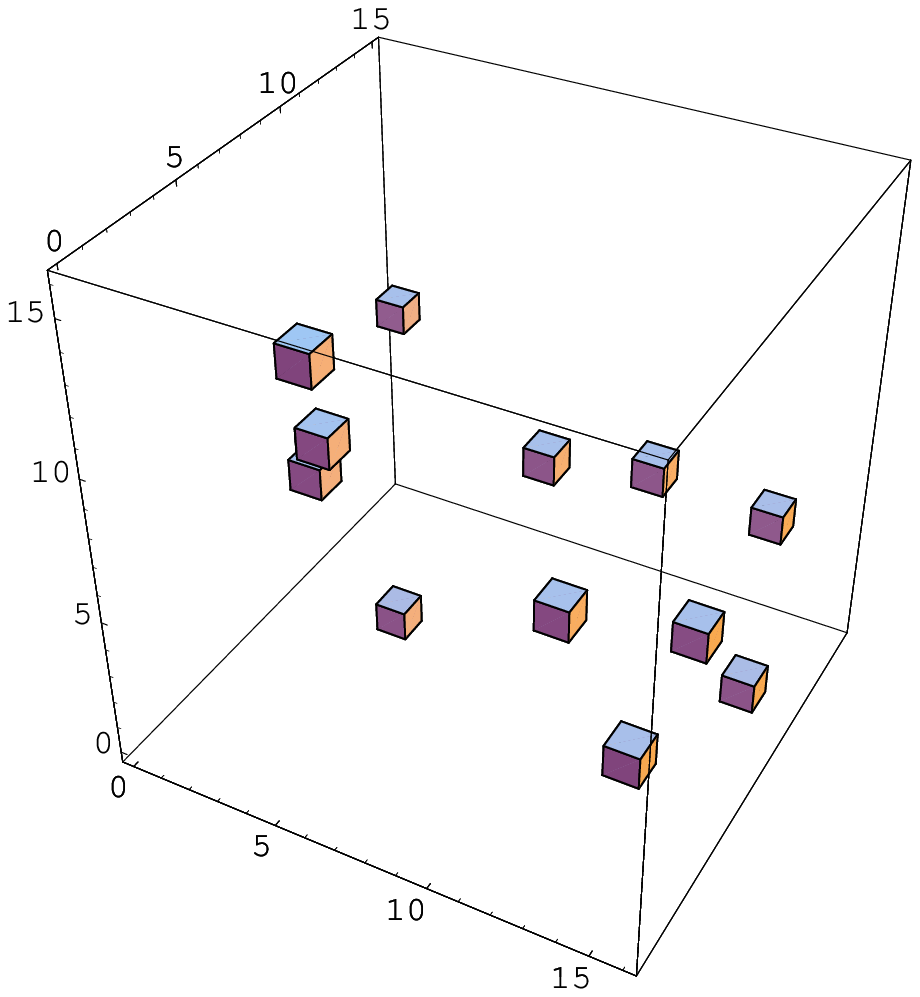,height=5cm}
\caption{Instanton  (left) and anti-instanton (right) 
vacuum configuration (MA gauge)}
\label{mono-conf-MA}
}

Polyakov assumed  that instanton charges $q_a=\pm 1$
alone contribute to the summation. 
We plot a histogram of numerically obtained instanton charges
in Fig.\ref{hist-ch} for various blocked instantons.
Theoretically an $n$ blocked instanton can take a charge up to
$3n^2-1$. However as seen from Fig.\ref{hist-ch}, the number of
instantons having a charge larger than one is very small.
This is consistent with Polyakov's assumption.
\FIGURE{
\epsfig{file=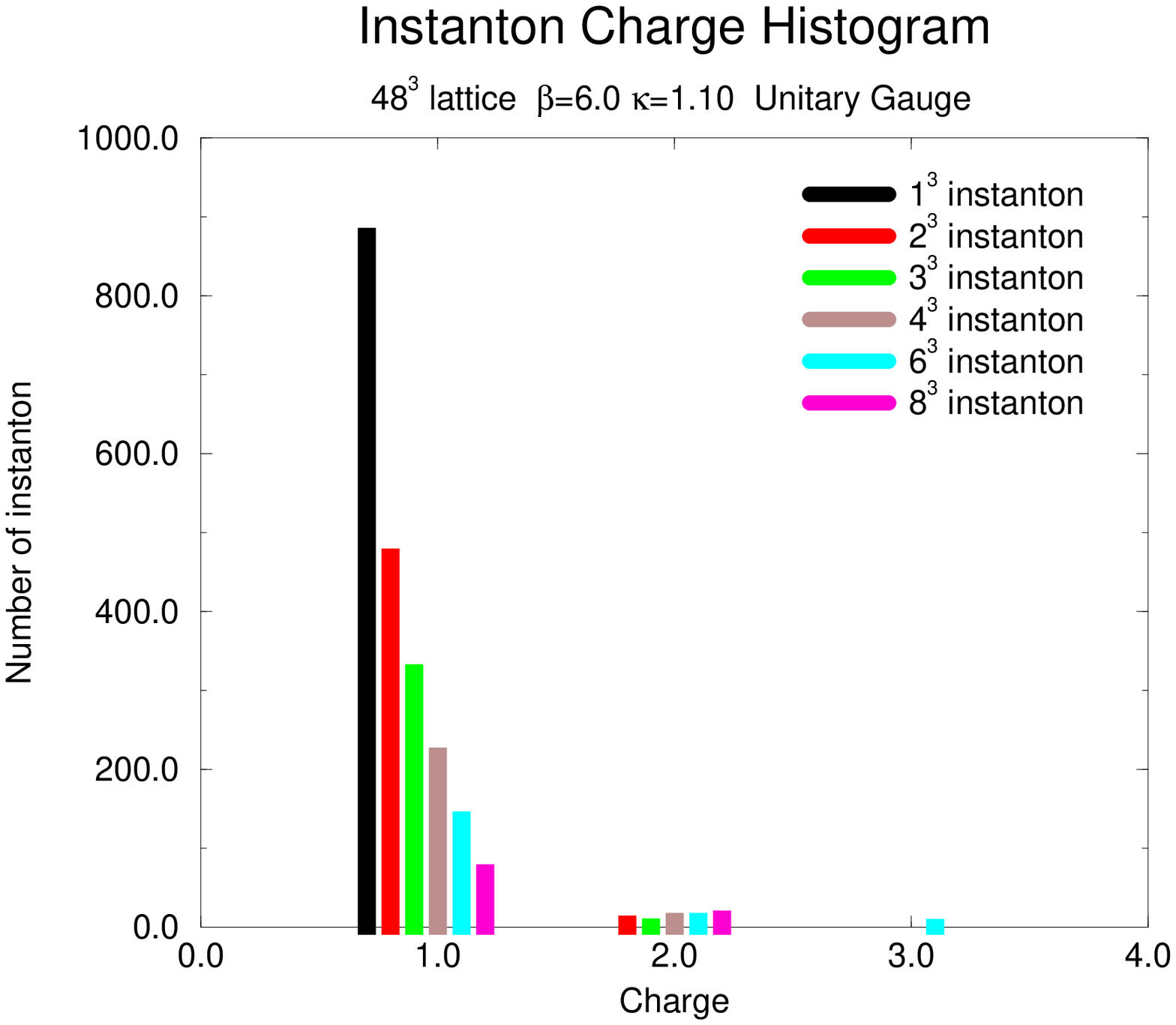,height=6cm}
\epsfig{file=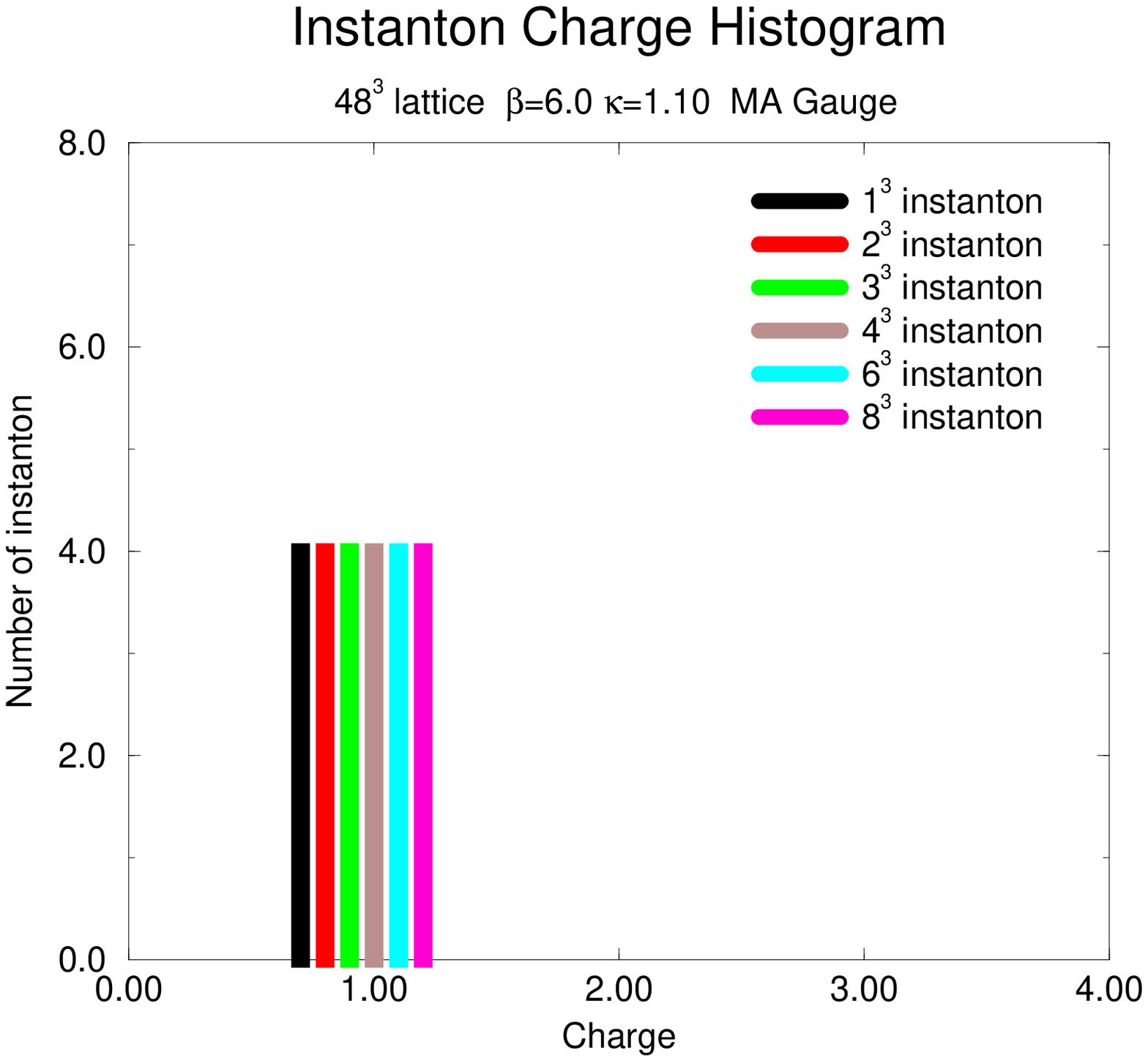,height=6cm}
\epsfig{file=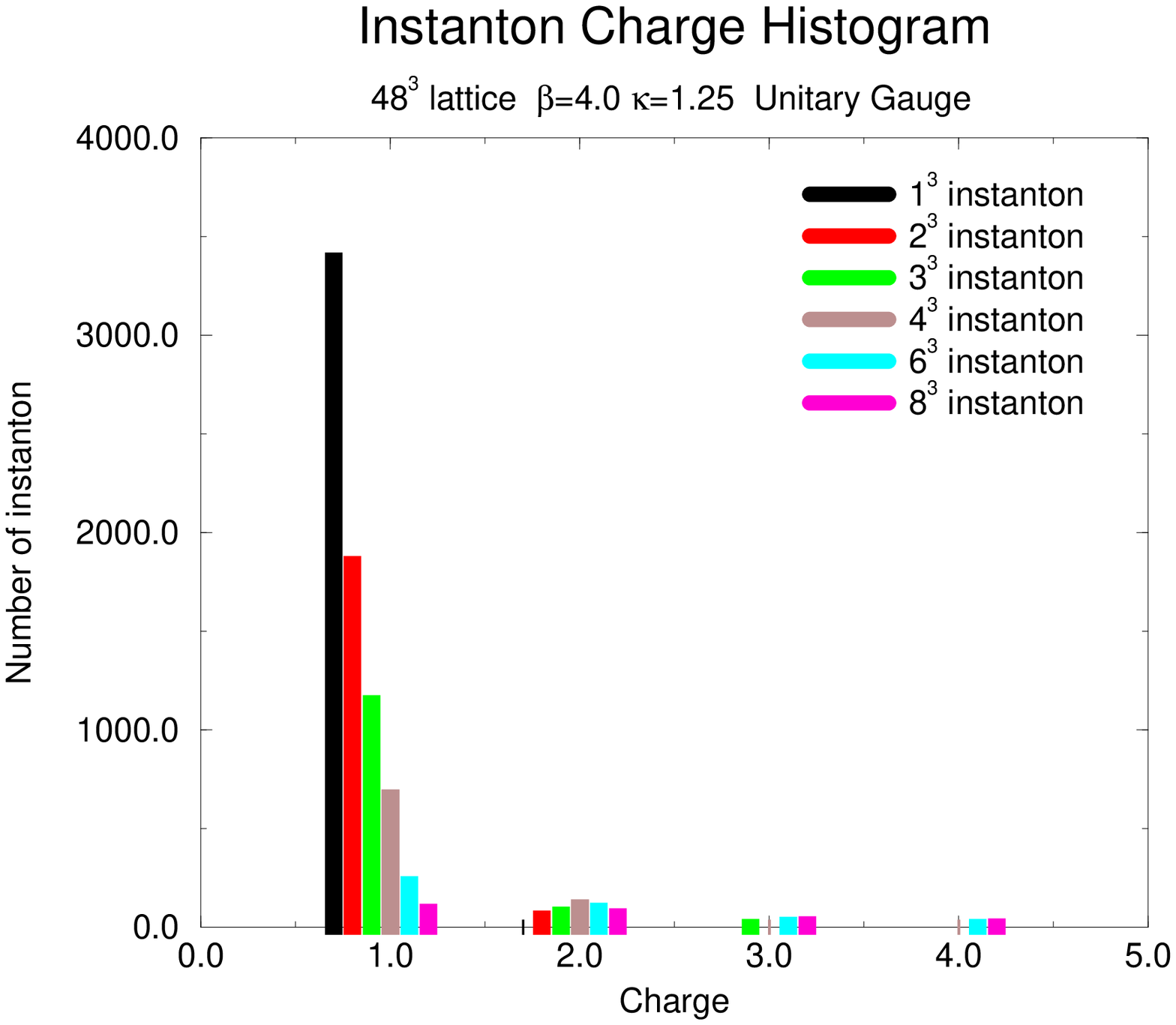,height=6cm}
\epsfig{file=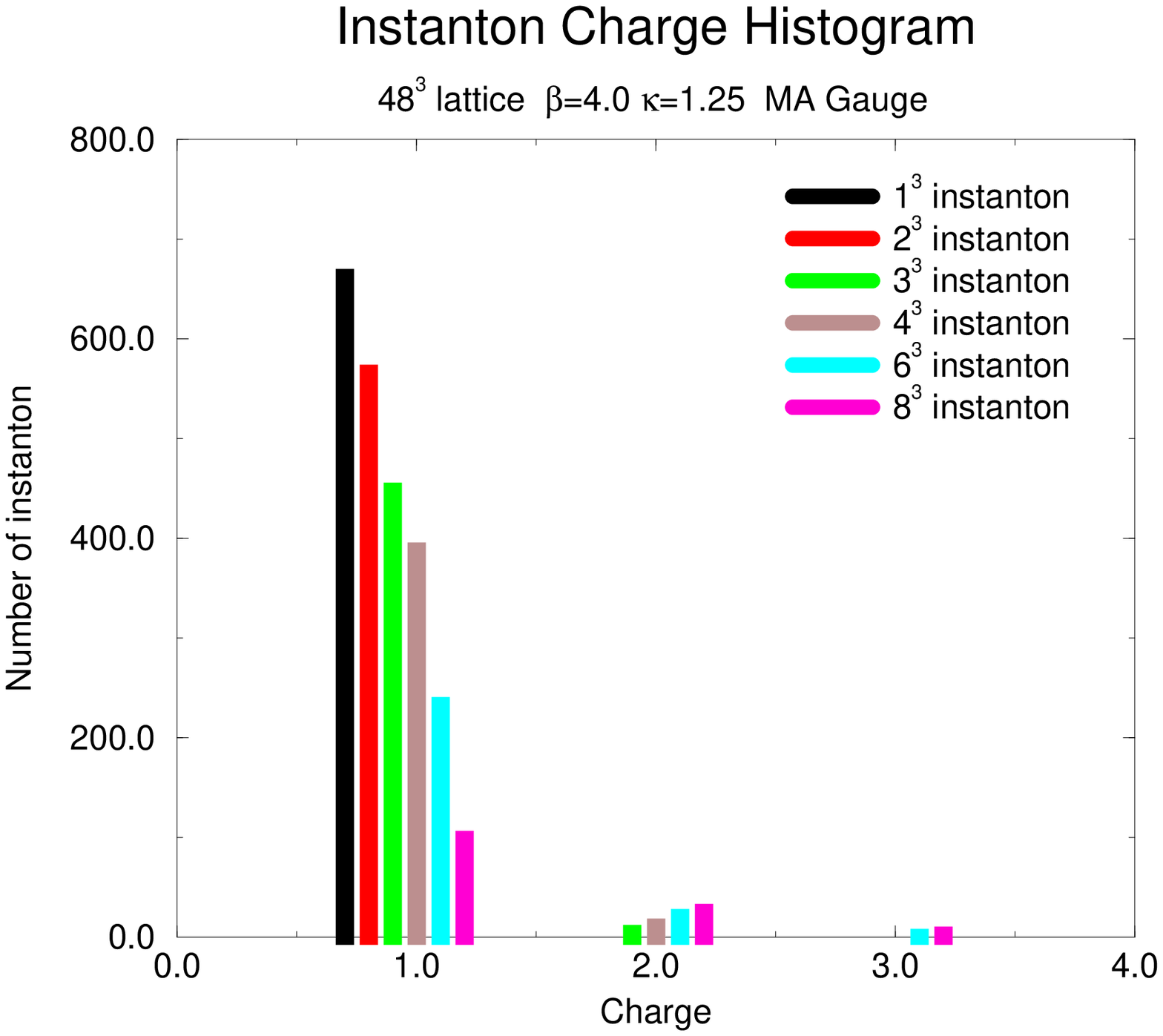,height=6cm}
\caption{Instanton charge histogram}
\label{hist-ch}
}

\subsection{Parameter regions of $\beta$ and $\kappa$ }
Let us discuss here what values of the parameters 
$\beta$ and $\kappa$ should be chosen.
We plot  the static potential in
Fig.\ref{k-dep-pot} for different values
of $\kappa$ with fixed $\beta=5.0$ 
and the instanton density in Fig.\ref{mono-dens} for 
$\beta=6.0$. 

We see that, 
in the case of the unitary gauge condition, 
the instanton density of the system is very large  for small $\kappa$, 
whereas in the case of MA gauge condition it remains low. 
The dilute gas approximation may be wrong for small $\kappa$ regions 
in unitary gauge. Actually, the abelian and the instanton 
potentials have large errors in the unitary gauge, whereas in MA gauge
these potentials are seen clearly.
The Georgi-Glashow model is reduced to  pure 
$SU(2)$  gauge theory in the limit of $\kappa\rightarrow 0$.
Such density and potential 
behaviors of both gauges are seen also in four-dimensional 
pure $SU(2)$  gauge theory\cite{Suzuki-rev}.
%
%

What happens in the large $\kappa$ region? 
One can find that there seems to be a
critical $\kappa_c$ above which 
the instanton density becomes vanishing and the instanton contribution
to the potential becomes zero.
Note that the Georgi-Glashow model  in three dimensions is always in 
the confinement phase contrary to the model in four dimensions.
The existence of $\kappa_c$ does not mean that 
instantons are not responsible for confinement.
In the limit of $\kappa\rightarrow\infty$,  the three-dimensional 
Georgi-Glashow model becomes the three-dimensional compact U(1) model.
The confinement in the latter model is proved to be due to instantons 
for all $\beta$ region\cite{Banks-Kogut}. 
We can define instantons in the compact U(1) model naturally without
following DeGrand and Toussaint (DT). It is those instantons which are
responsible for confinement in the compact U(1) model. (We call it as natural 
instanton.)
This concludes that the DT definition of instanton 
is not good for larger $\kappa$ region than $\kappa_c$ 
in three-dimensional Georgi-Glashow model.
In Ref.\cite{Shiba-suzuki2}, the effective actions of DT 
and natural monopoles are studied in 4D compact QED.
They are not always in agreement. Hence it is not understandable
 that the difference 
appears for such large $\kappa$ and $\beta$ region also in 3D (GG)$_3$.
We admit that the physical meaning of $\kappa_c$ is not understood 
theoretically, however.
In this numerical study, we have to adopt  
the DeGrand-Toussaint definition of instanton, 
since we do not know 
how to define natural monopoles in 3D (GG)$_3$.

Instantons in
the unitary gauge are more interesting, since they correspond to the 
instantons studied by Polyakov\cite{Polyakov77}. Hence 
 in the following discussions we restrict ourselves to  
the value of  $\kappa$  around $1.0$ which is near  but less
than $\kappa_c$. 

\FIGURE{
\epsfig{file=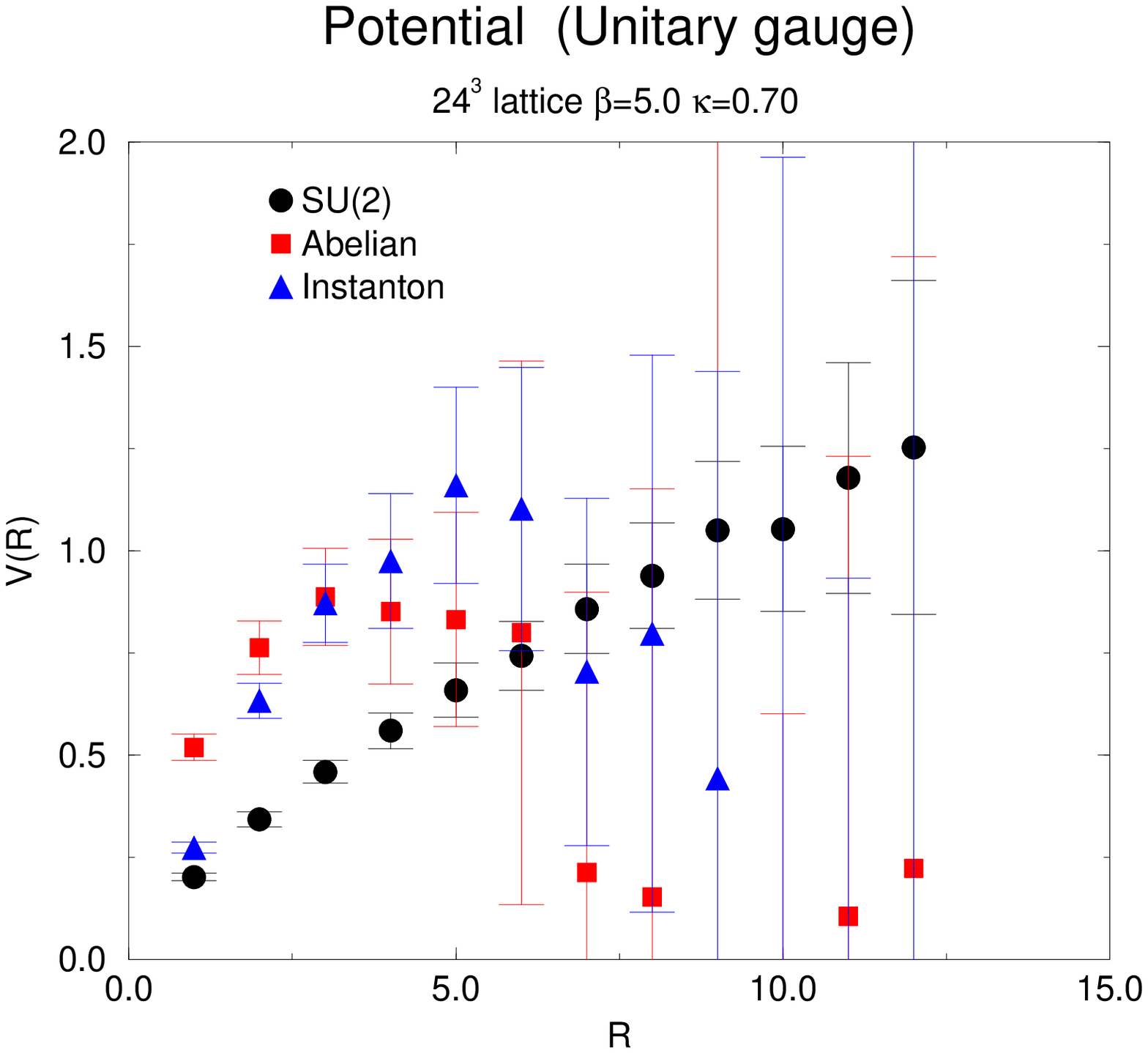,height=6cm}
\epsfig{file=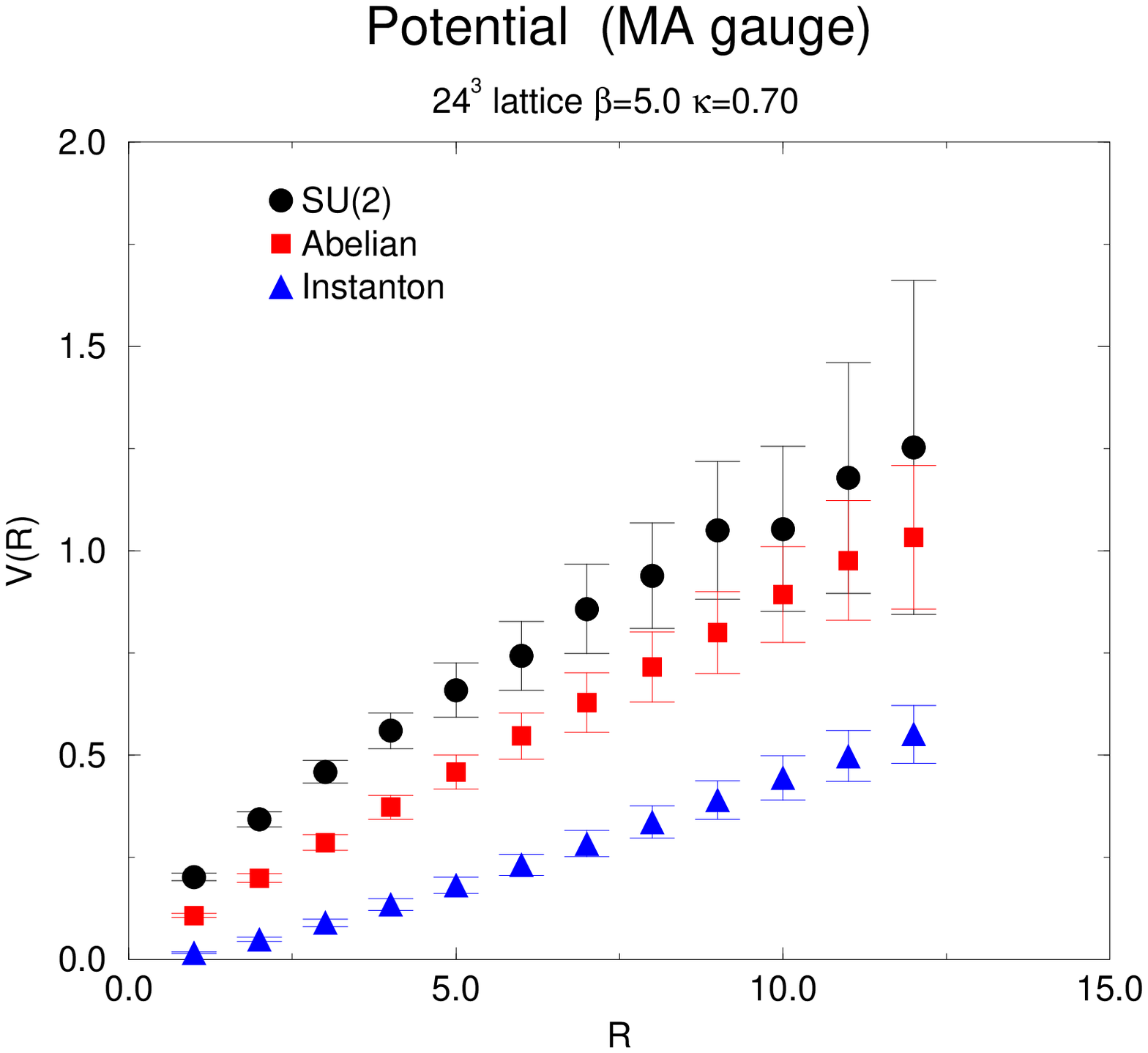,height=6cm}
\epsfig{file=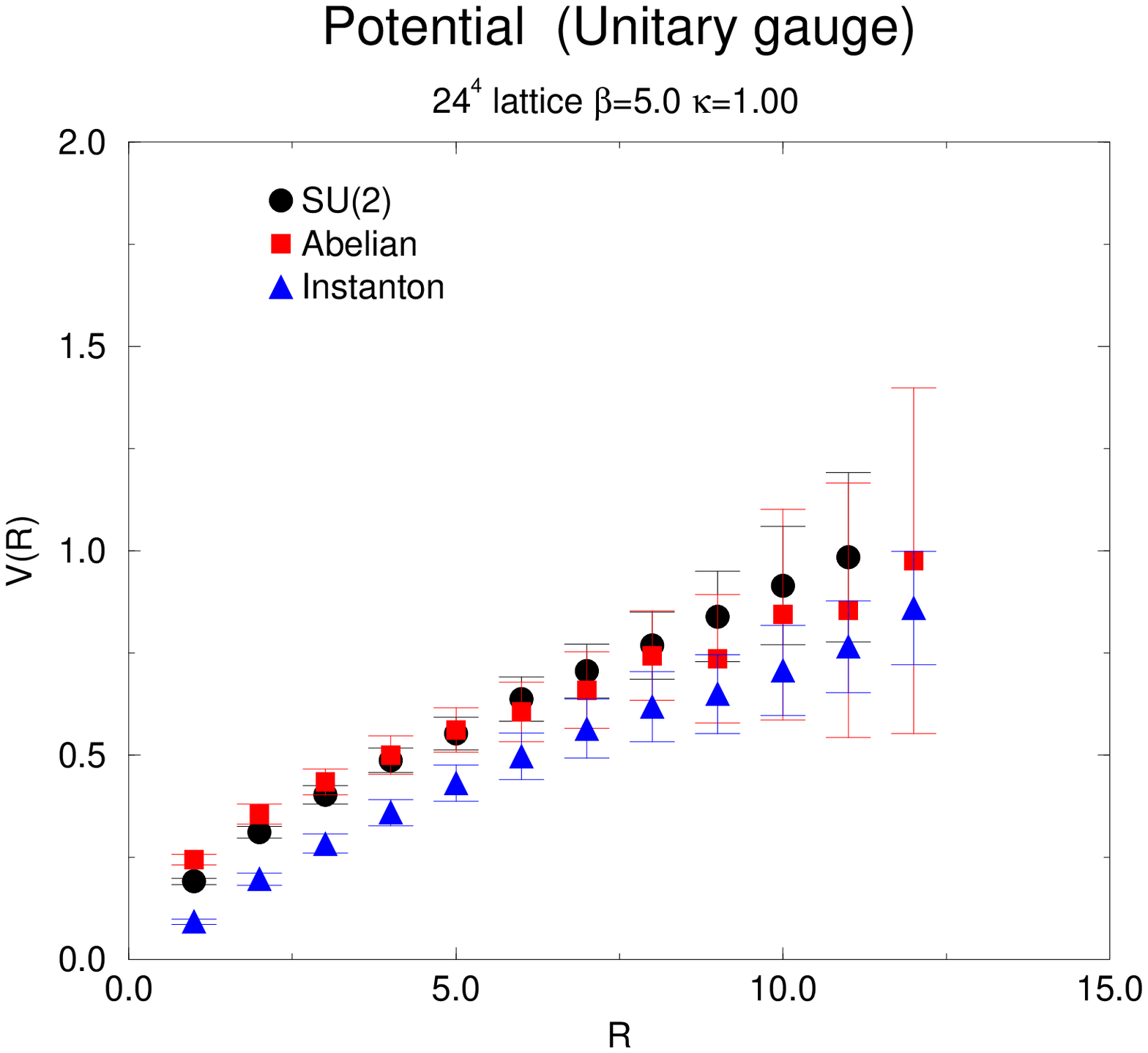,height=6cm}
\epsfig{file=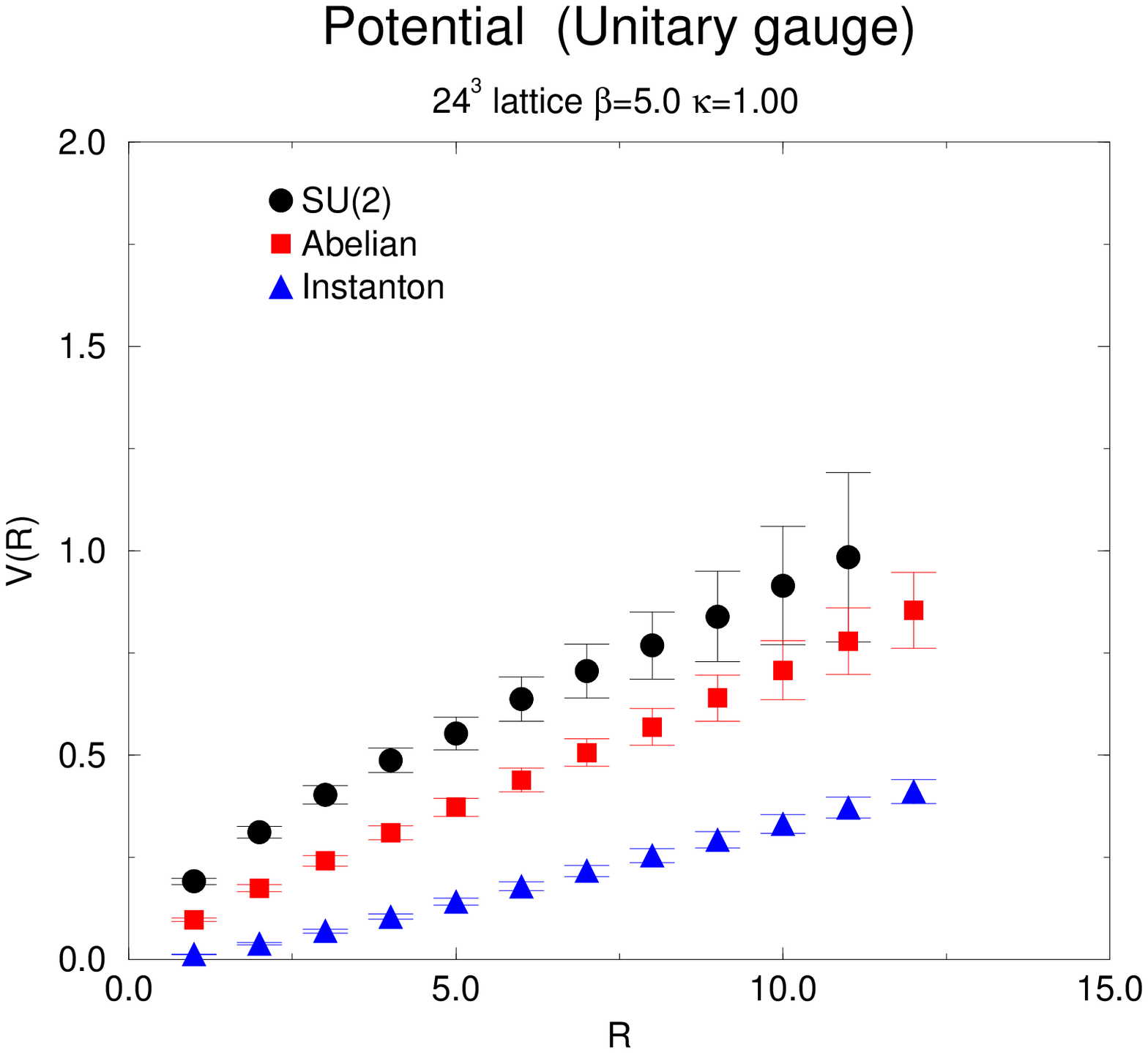,height=6cm}
\epsfig{file=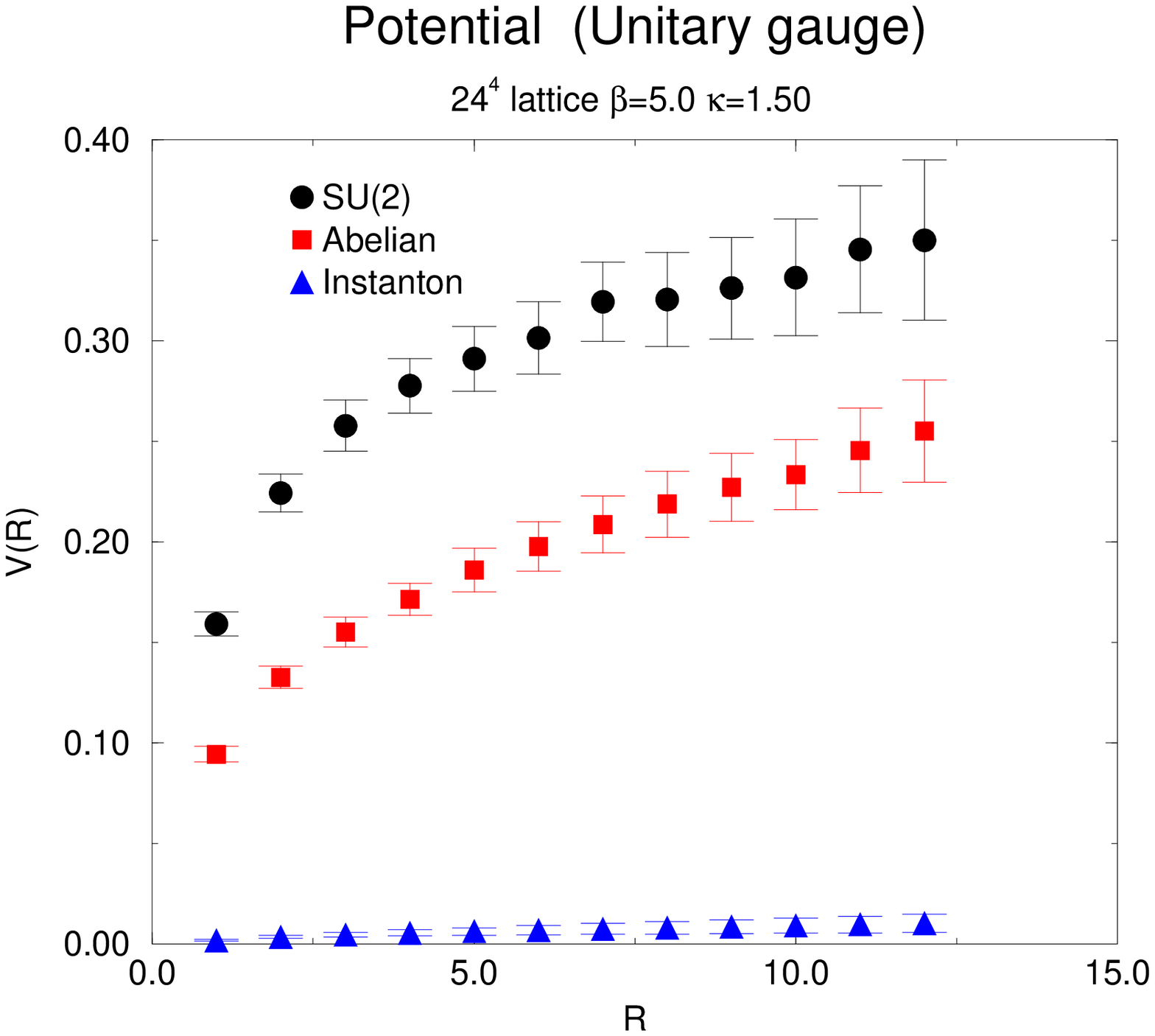,height=6cm}
\epsfig{file=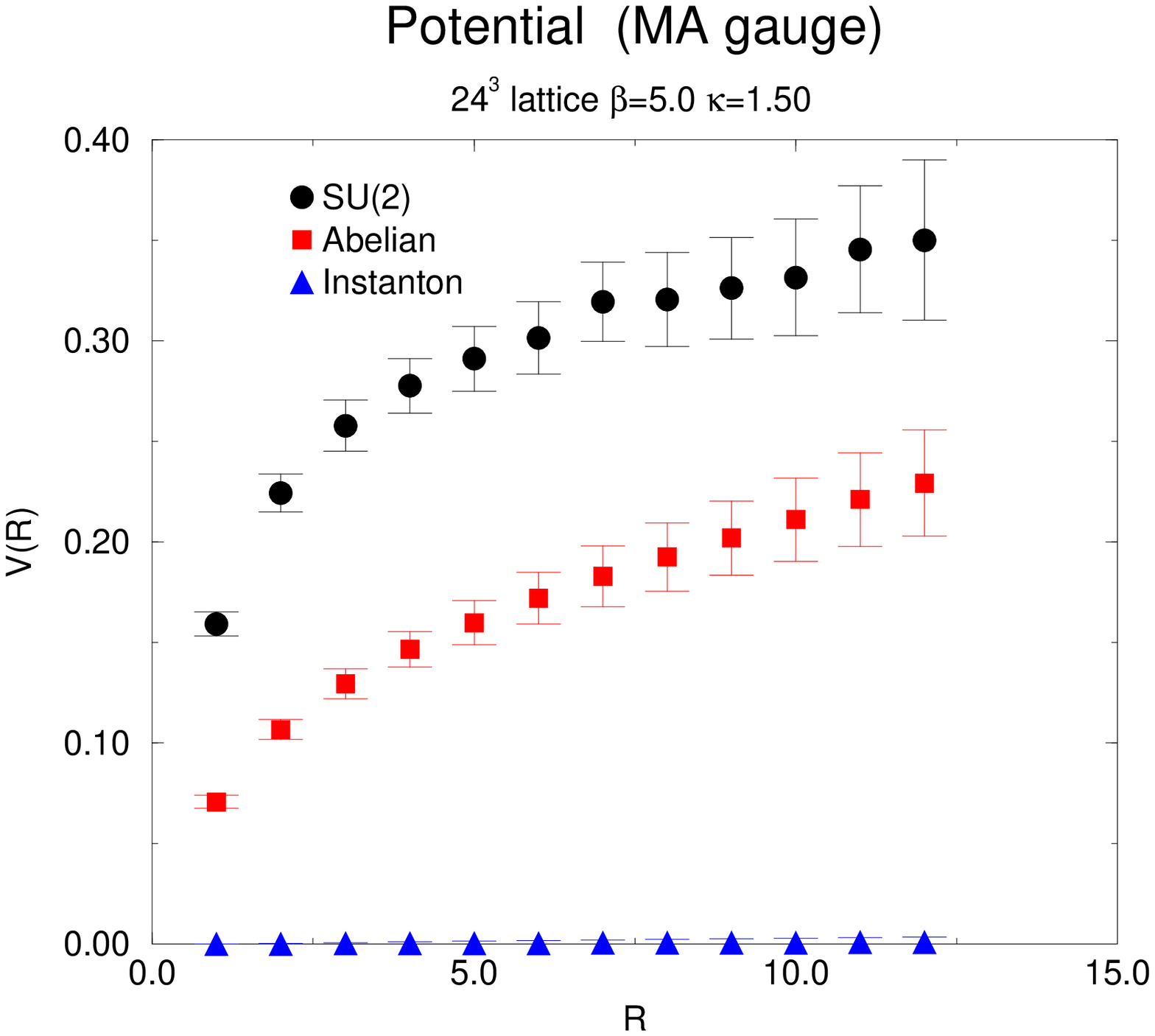,height=6cm}
\caption{$\kappa$ dependence of $q$-$\bar{q}$ potential}
\label{k-dep-pot}
}
\FIGURE{
\epsfig{file=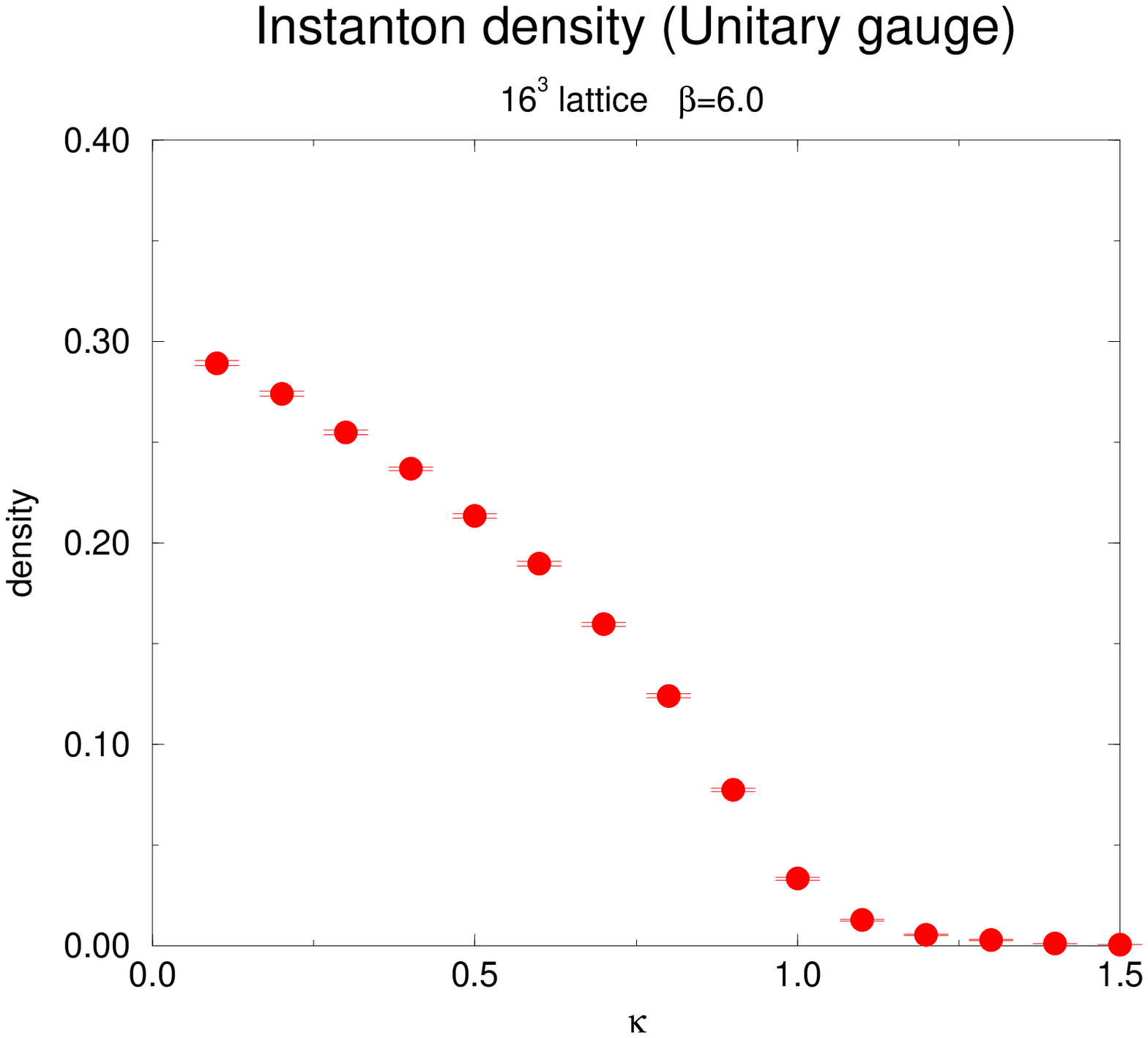,height=5cm}
\epsfig{file=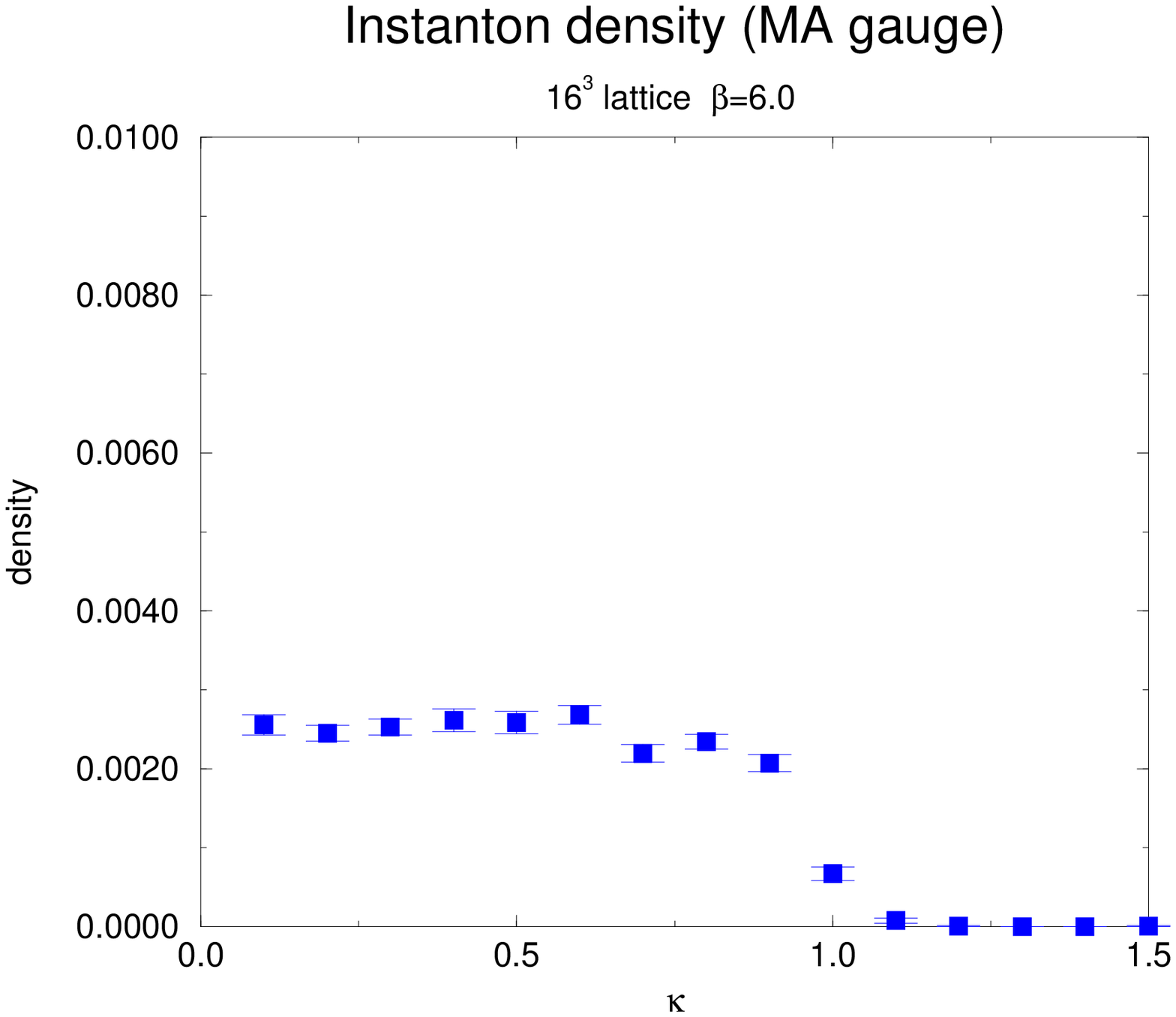,height=5cm}
\caption{Instanton density}
\label{mono-dens}
}

What about the $\beta$ dependence?
We show in Fig.\ref{b-dep-pot} the $\beta$ dependence for fixed 
$\kappa=1.00$. 
For small $\beta$, the static potential for large distance is not fixed
clearly. This comes from that the theory in 
the small $\beta$ region has a large lattice 
string tension which corresponds to a large lattice distance as shown
later. To study the continuum limit, it is better to study larger $\beta$
region.

However, for very large $\beta$,  even $\kappa=1.00$ becomes larger than
$\kappa_c$ in MA
gauge. Namely $\kappa_c$ depends on the value of $\beta$. When 
$\beta$ becomes larger, $\kappa_c$ becomes smaller. $\kappa_c$ in MA
gauge is a little bit smaller than that in the unitary gauge.
Since we discuss around $\kappa=1.00$, we are restricted to 
$\beta=4\sim 7$ regions.

\FIGURE{
\epsfig{file=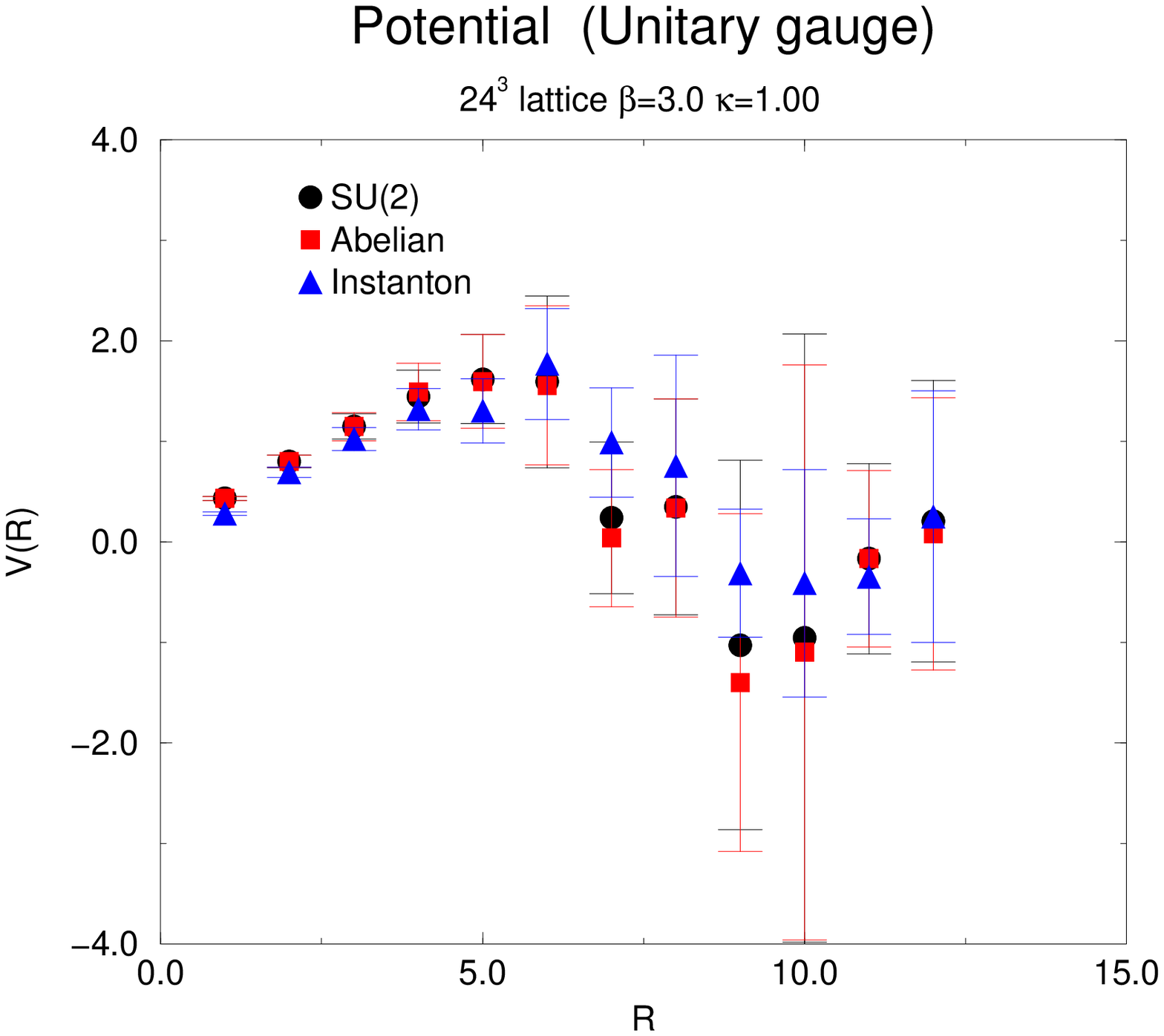,height=6cm}
\epsfig{file=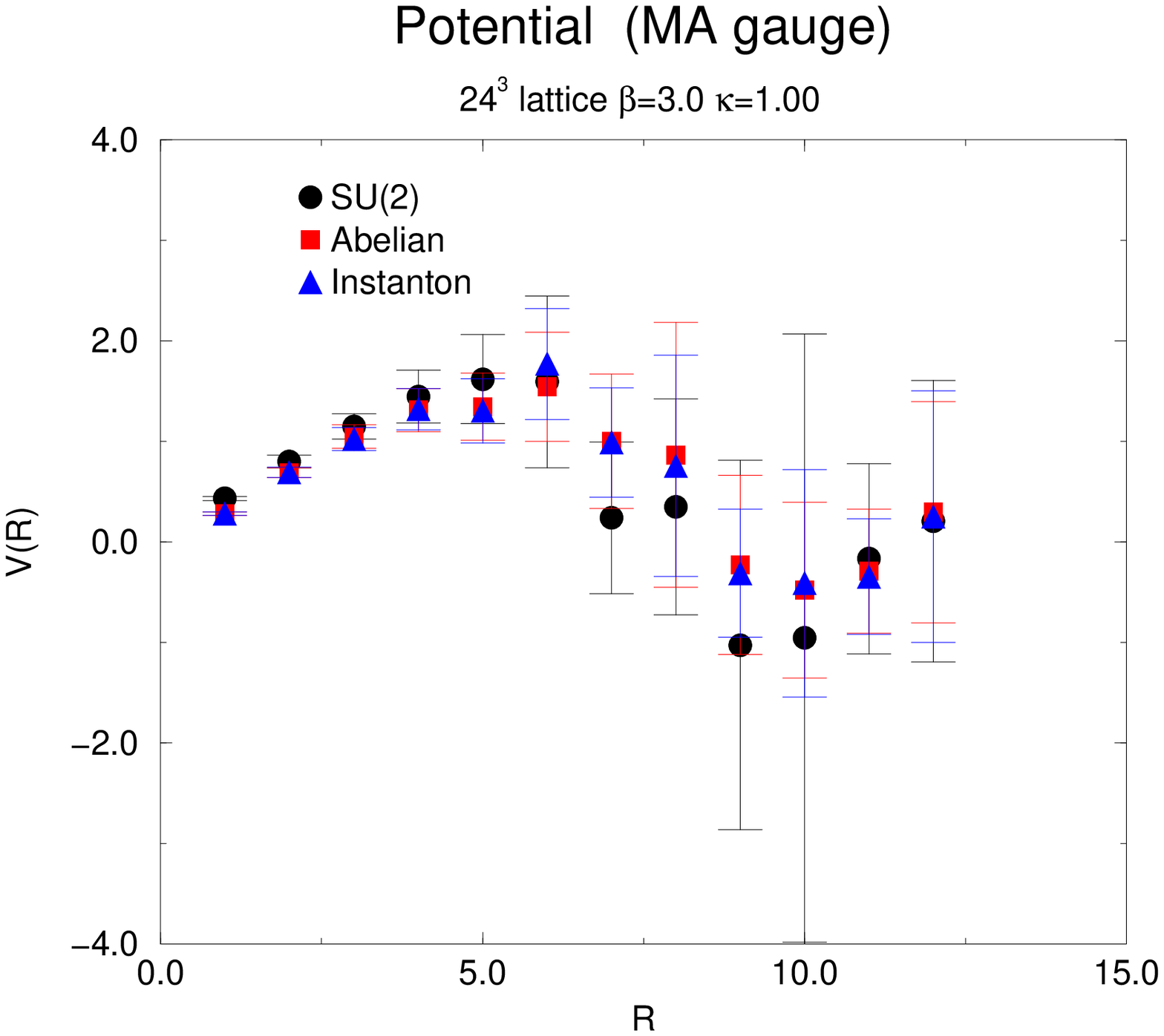,height=6cm}
\epsfig{file=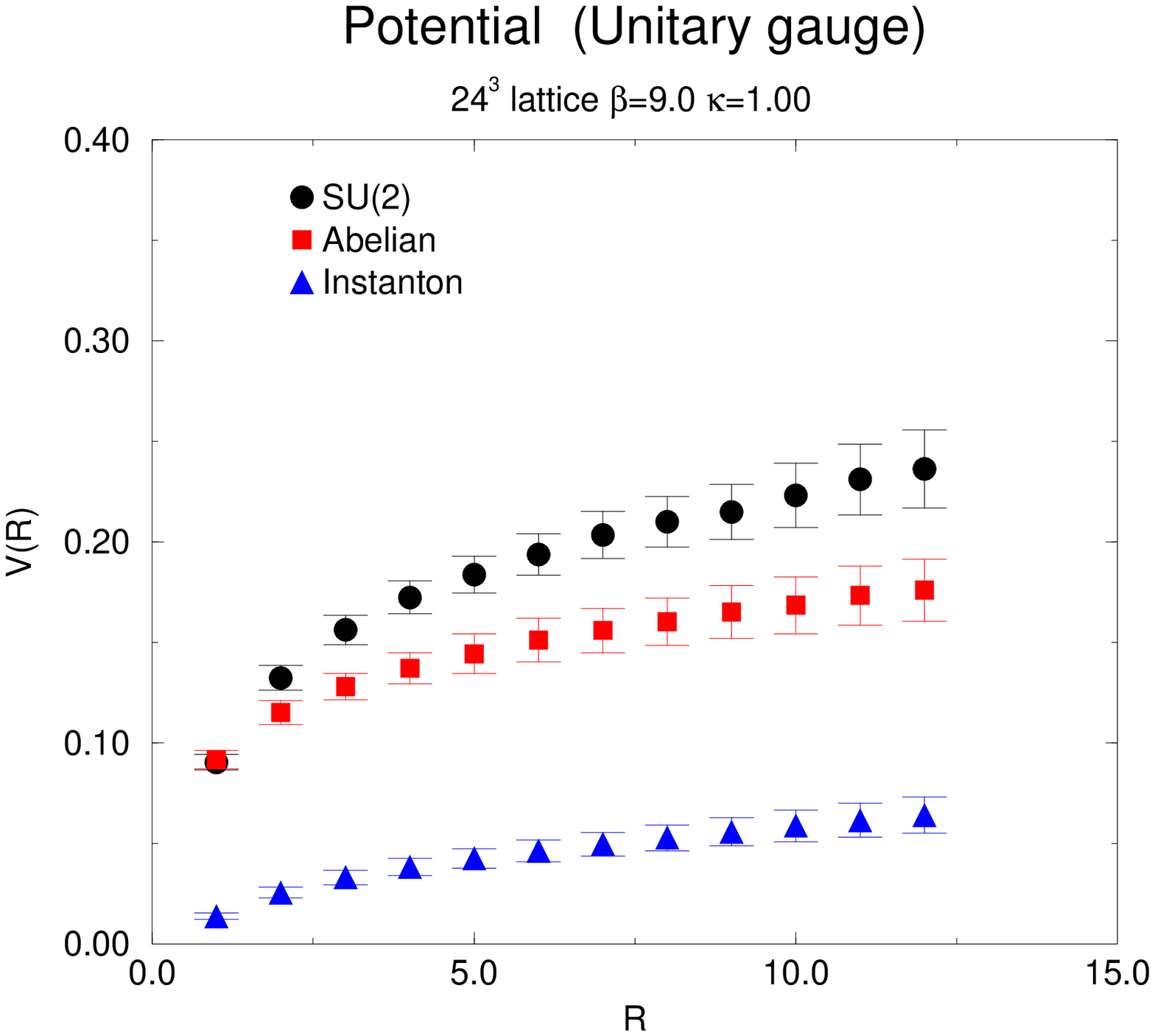,height=6cm}
\epsfig{file=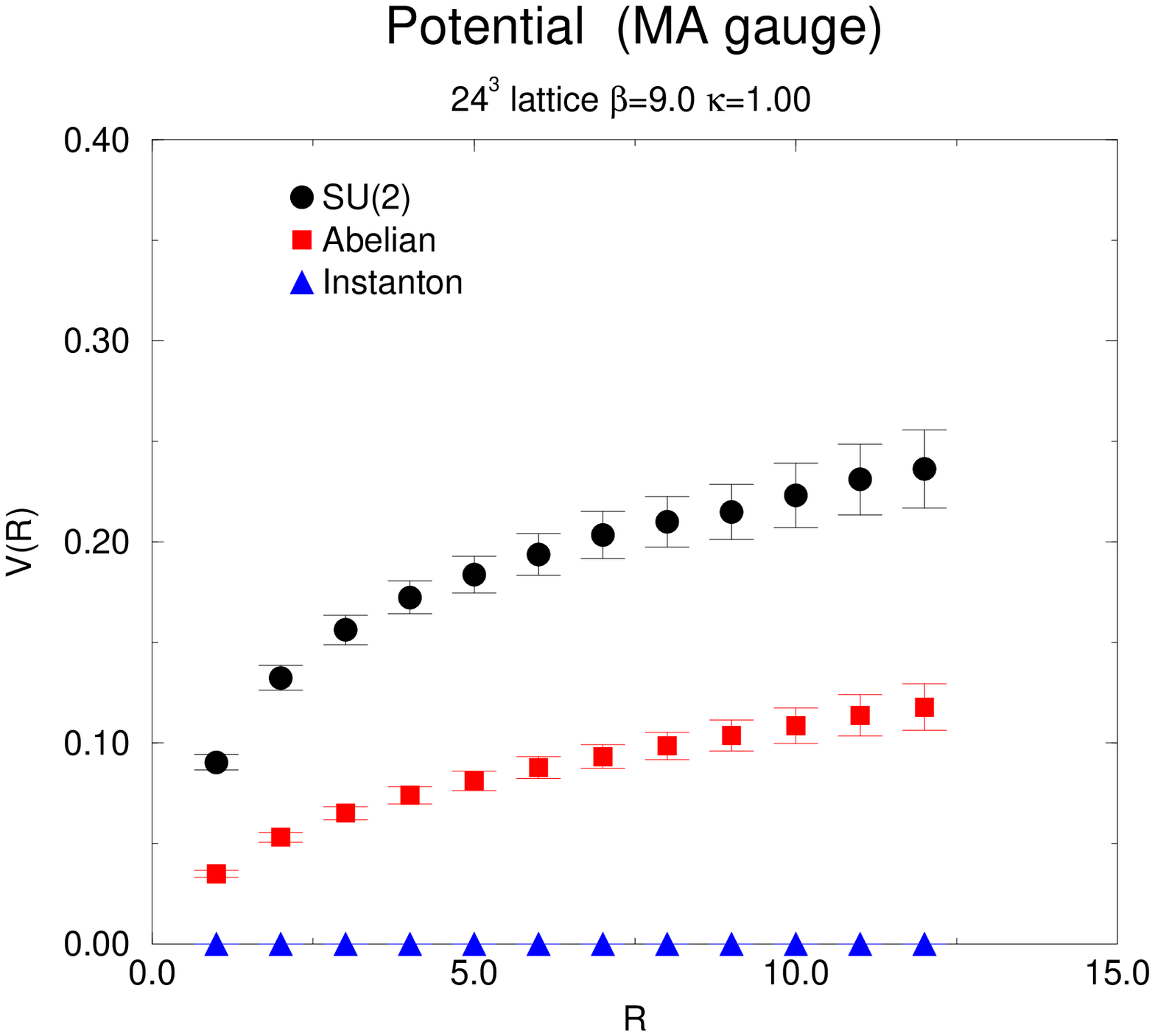,height=6cm}
\caption{$\beta$ dependence of $q$-$\bar{q}$ potential}
\label{b-dep-pot}
}

\subsection{Lattice instanton action}

Since the instanton dominance is seen numerically, an effective
instanton action is expected to work in the infrared region.
Let us  derive the lattice instanton action from instanton configurations
using Swendsen's method \cite{Swendsen} (See also Appendix \ref{swend}).
We express $D(x)$ in (\ref{lat-instanton-action}) as
\begin{equation}
D(x-x')=G_0\delta^3(x-x')+G_1(\delta^3(x - (x'-\hat{1}))+...) + G_2... ,
\end{equation}
where we restrict ourselves to ten different two-point couplings shown in 
Fig.\ref{action-operator}. 
%
%
Our numerical results are the following:
\begin{enumerate}
 \item 
The inverse Monte-Carlo method works very well and the  10 coupling
constants are fixed very beautifully. The convergence is very fast.

When we start from the ensemble of blocked instanton configurations 
$\{k^{n}(x)\}$, we can get also a lattice action $S[k^{n}]$ very
       rapidly.

We have tried to include a four-point self interaction in addition. In
       the case of $n=1$, we can fix the coupling constants, although 
the convergence is very slow. The convergence was not obtained in the
       case $n > 1$. Hence we take into account only the two-point
       interactions.
 \item 
We show in Fig.\ref{instanton-action-fig} an example of coupling
       constants in both gauges, where the horizontal axis stands for
       the distance between two instantons in each two-point 
interaction.  The data points of $S_8$ and $S_9$ in
       Fig.\ref{instanton-action-fig}
are almost degenerate.

The action (\ref{polyakov-model-action2}) 
derived by Polyakov\cite{Polyakov77} is composed of the self-interaction
       and the Coulomb term. Hence let us check if the above action 
obtained numerically can be written similarly as 
in (\ref{polyakov-model-action2}).  When we rewrite the lattice Coulomb
       propagator as 
\begin{equation}
\Delta_L^{-1}(x-x')
 =C_0\delta^3(x-x')+C_1(\delta^3(x - (x'-\hat{1}))+...) + C_2... ,
\label{coeff-coulomb}
\end{equation}
we get a beautiful fit  
\begin{equation}
{ G_i } \sim {\rm Const.} \times { C_i } ~(i \neq 0)
\label{G_i-C_i-fit}
\end{equation}
in both gauges as shown in Fig.\ref{instanton-action-fig}.

The lattice instanton action of Georgi-Glashow model 
in three-dimension can be  written as
\begin{equation}
S={\rm Self ~ term } + {\rm Const.}\times
  \sum_{x,x'}k(x)\Delta_L^{-1}(x-x')k(x').
\label{lat-a_Coulomb}
\end{equation}
Although the Coulomb term is reproduced well, we can not compare the
       self terms of the lattice and the Polyakov actions, since 
we need to know the details of short-ranged renormalization effects. 
Actually the lattice Coulomb propagator is finite even at zero distance,
       whereas the propagator in the continuum is infinite.
\FIGURE{
\epsfig{file=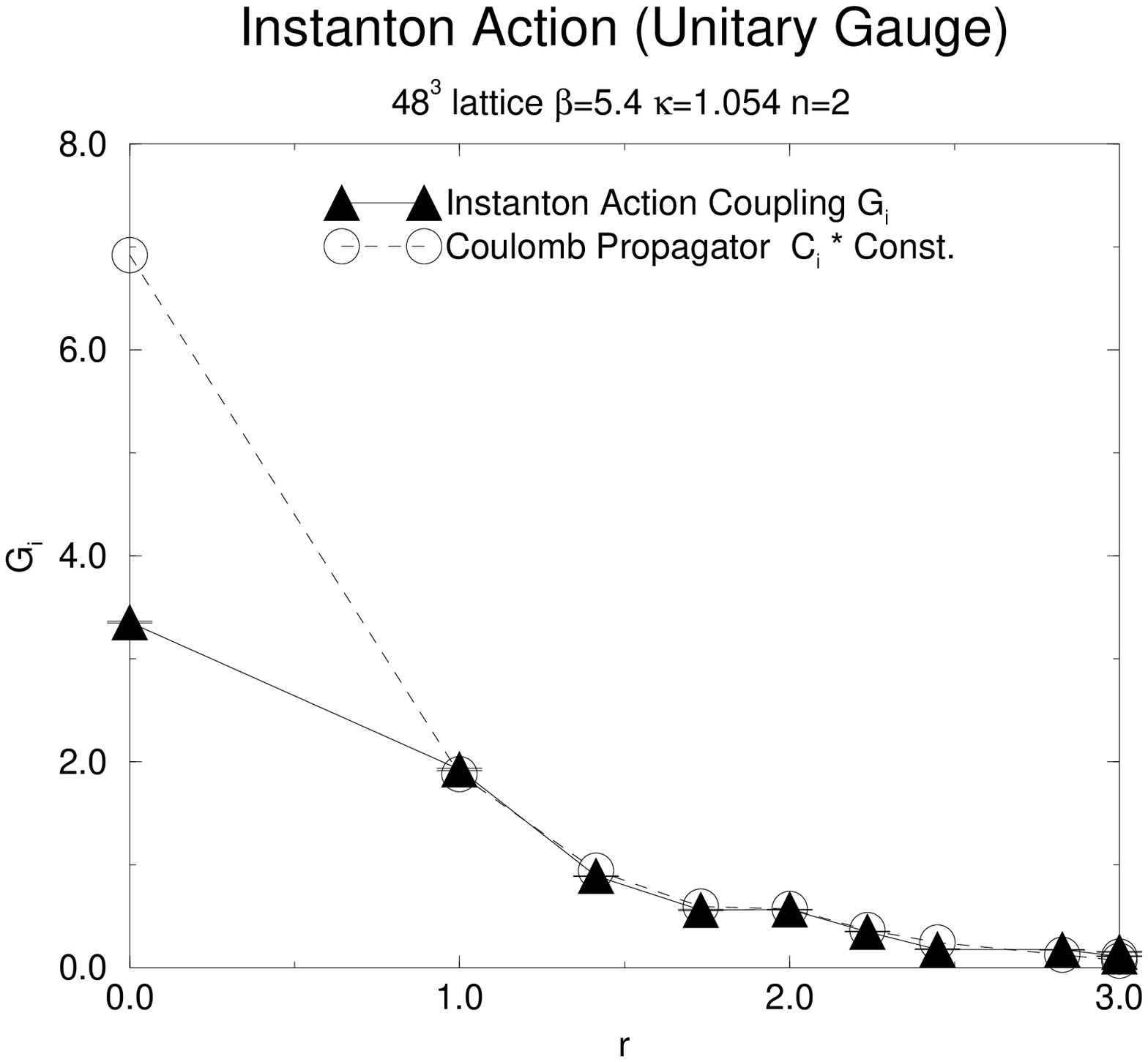,height=6cm}
\epsfig{file=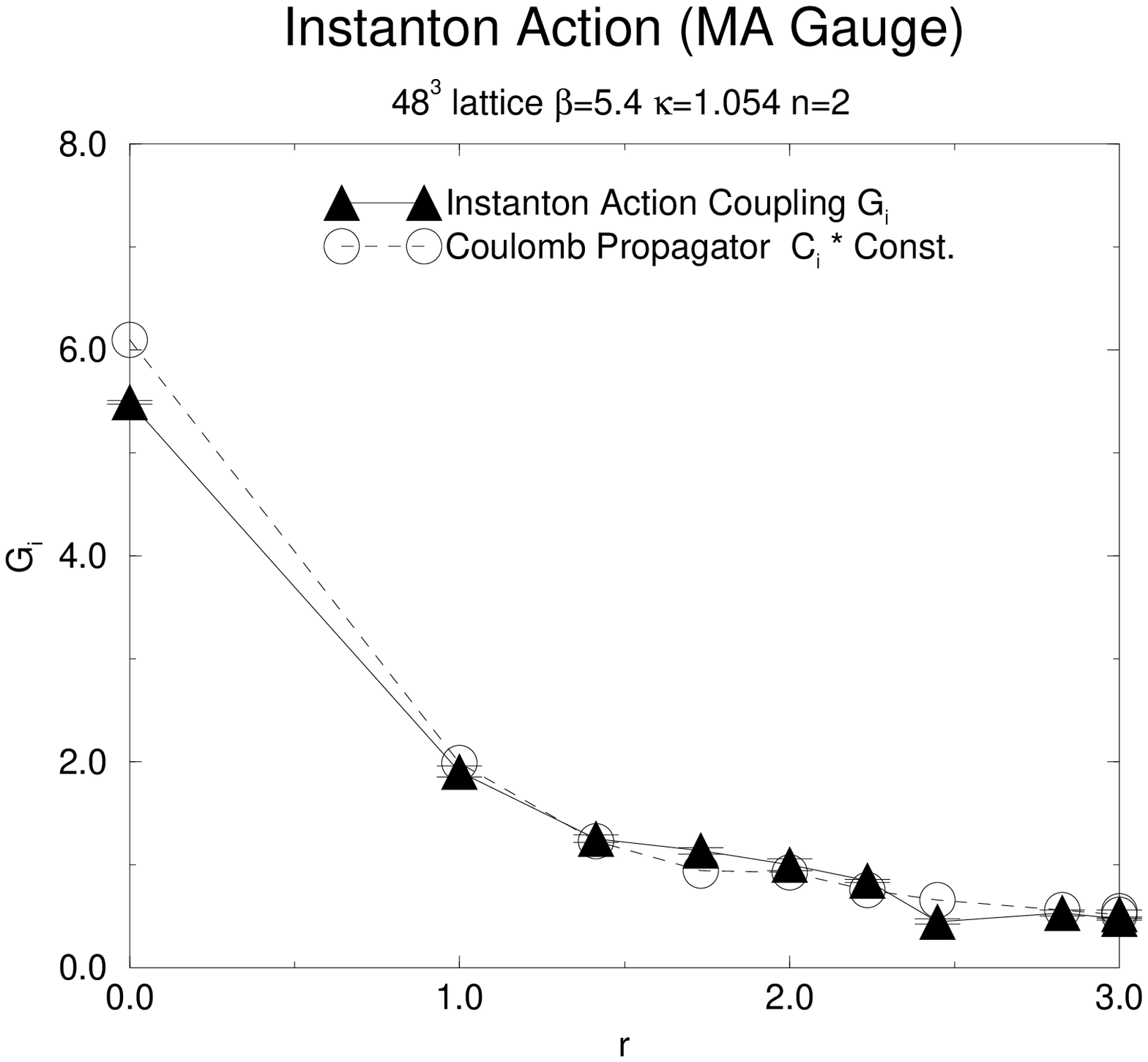,height=6cm}
\caption{Coupling constant of lattice instanton action and Coulomb propagator}
\label{instanton-action-fig}
}
 \item 
To study the continuum limit, we need to take both $V\to\infty$ and 
$a\to 0$ limits.
First we study the volume dependence of the action, adopting 
$12^3, 16^3, 24^3, 32^3$ and $48^3$ lattices.
Fig.\ref{vol} is an example of 
the lattice volume dependence of couplings for $\beta=4.0, \kappa=1.225$
and $n=2 $.
In both gauge conditions, 
the lattice instanton action is almost independent of the  lattice volume.
\FIGURE{
\epsfig{file=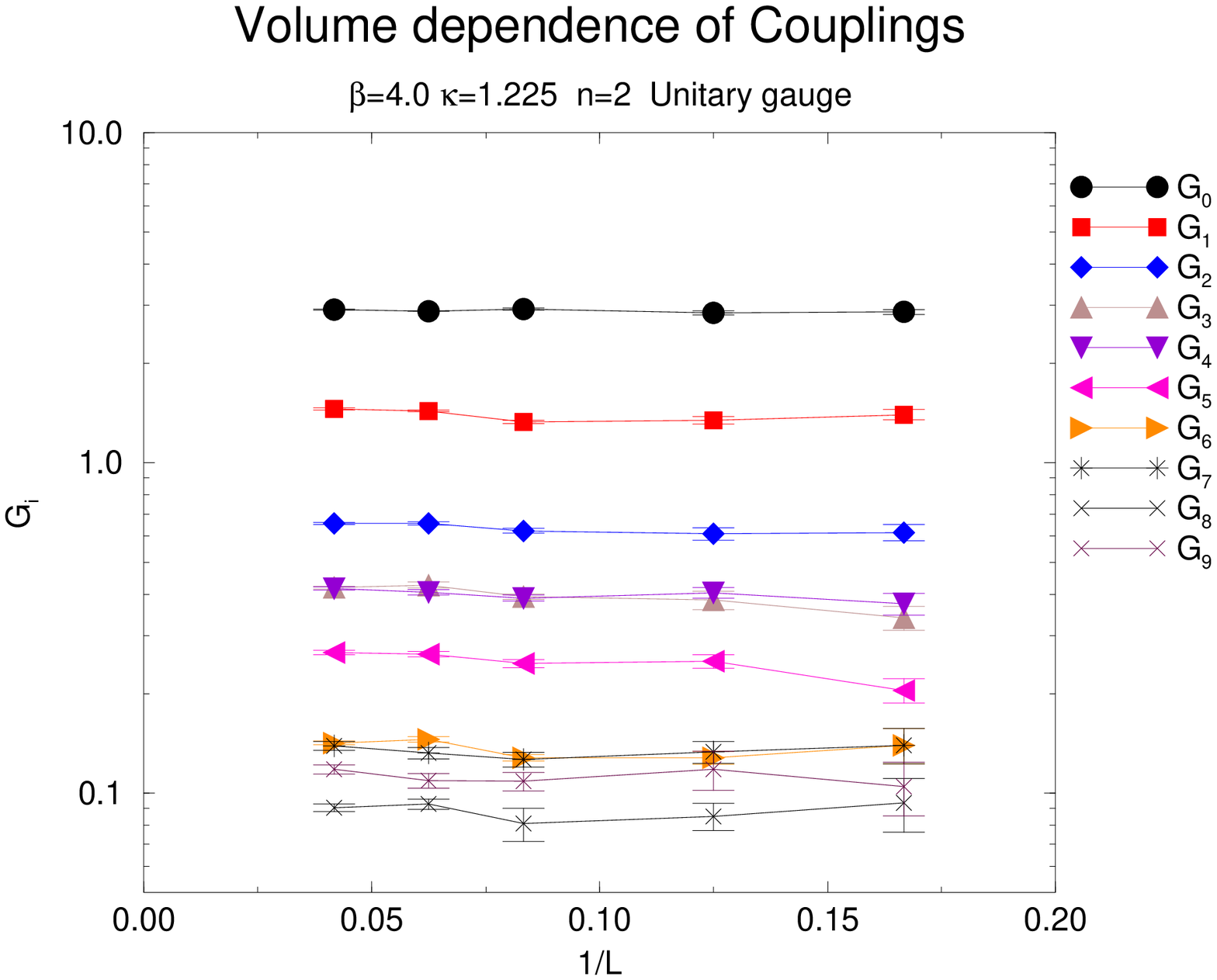,height=5cm}
\epsfig{file=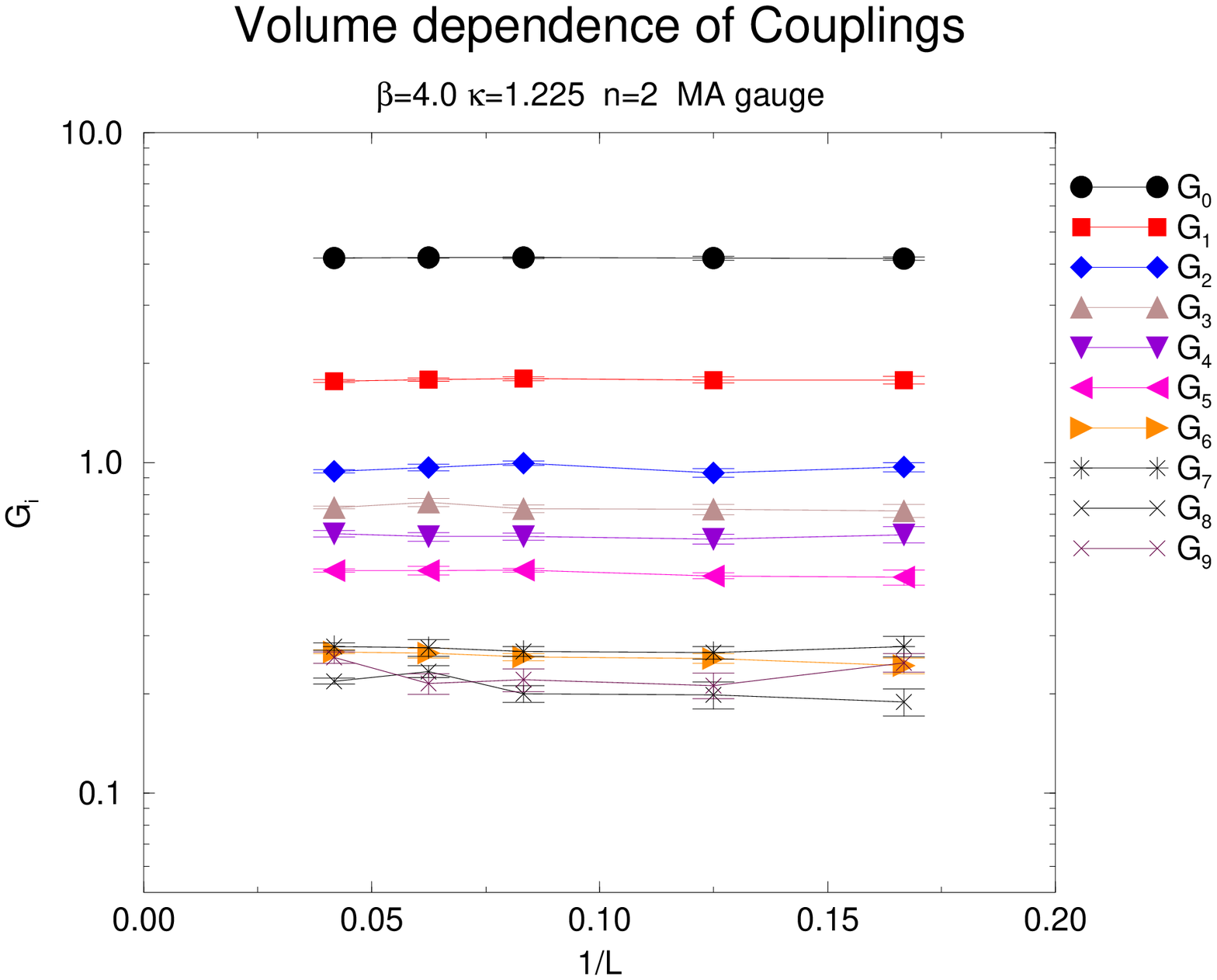,height=5cm}
\caption{Volume dependence of Coupling constants}
\label{vol}
}
 \item 
We next study the $a\to 0$ limit, performing the block-spin
       transformation (\ref{extend-instanton}). 
To get the renormalization flow, we have to fix a scale unit.
Since there is no physical experimental data 
in three-dimensional Georgi-Glashow model,
we fix the scale in terms of  the string tension determined 
from lattice simulations.
The lattice spacing is defined with
\begin{equation}
a = \sqrt{\frac{\sigma_L}{\sigma_{\rm phys}}},
\label{lattice-unit}
\end{equation}

We perform the block-spin transformations using the scale.
An example of the renormalization flow is plotted on the 
$G_1(b)-G_2(b)$ and the $G_2(b)-G_3(b)$ projected planes in Fig.\ref{g1-g2}.
We see that there seems to exist a universal flow line especially 
in the infrared large $b=na(\beta,\kappa)$ region.  
Each flow line for large $b$ is straight with small errors.
The universal flow lines of both gauges are almost the same for large $b$.
This suggests the gauge independence of the action in the infrared 
regions of the continuum limit.
\FIGURE{
\epsfig{file=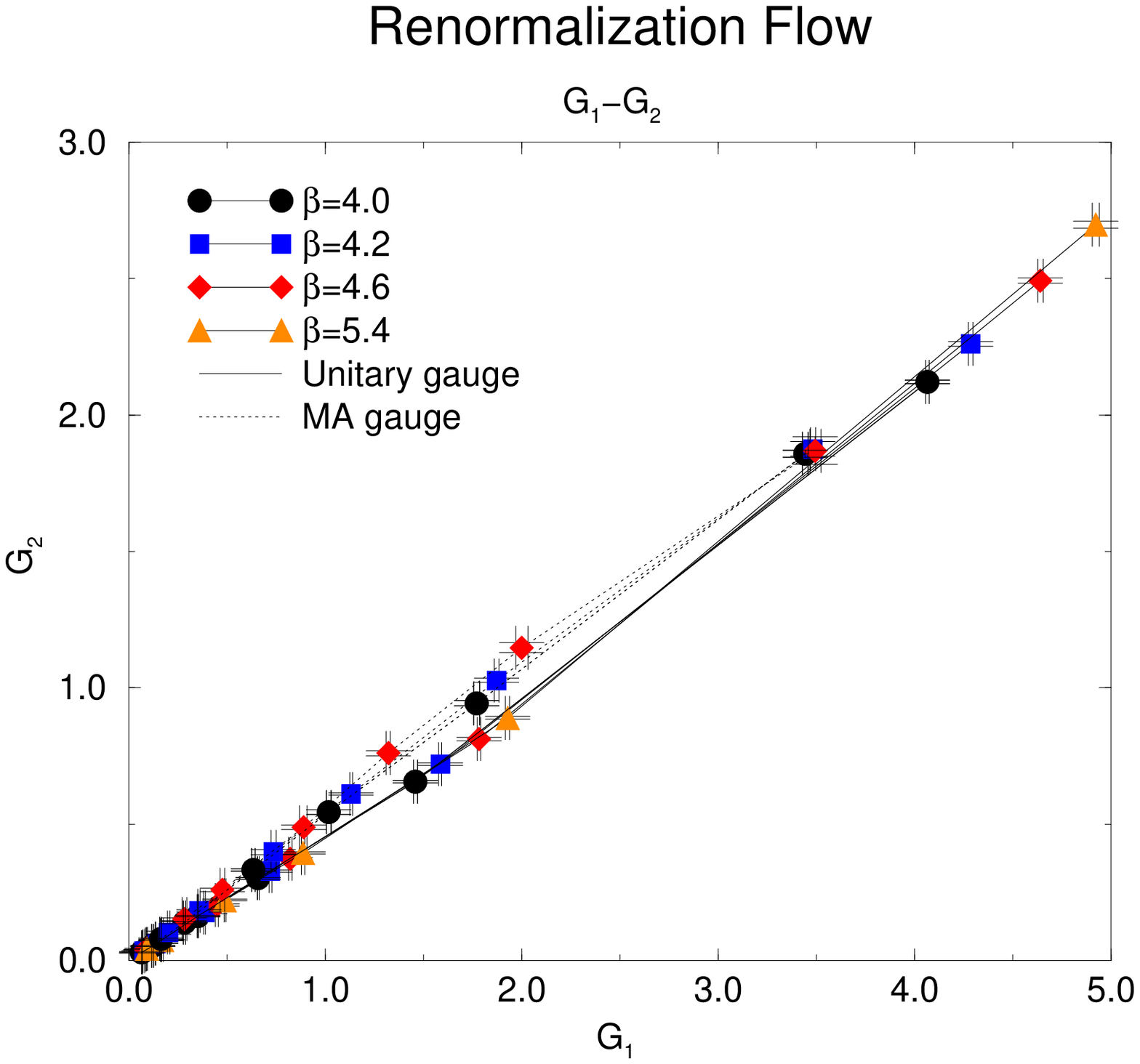,height=5.5cm}
\epsfig{file=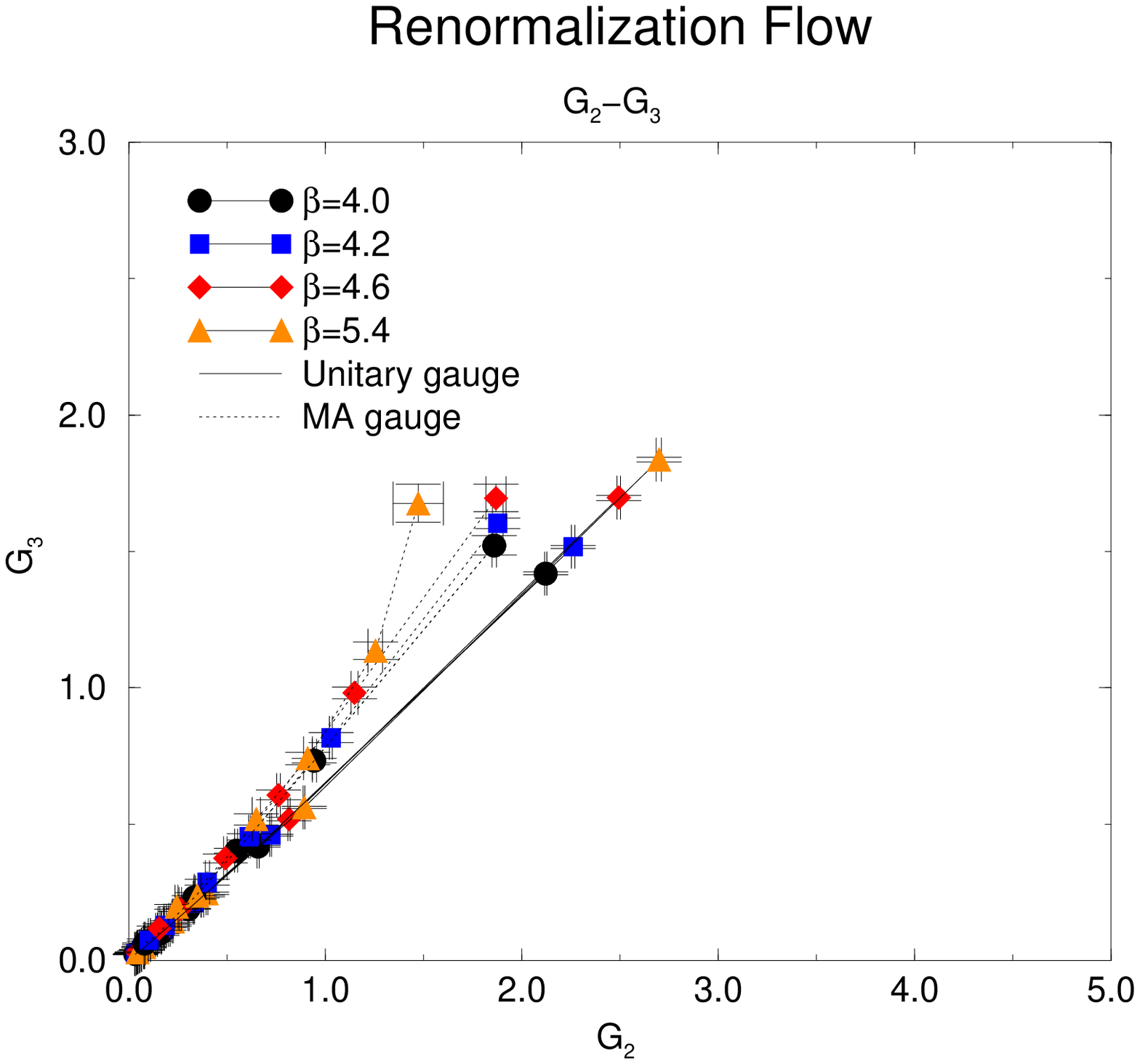,height=5.5cm}
\caption{Flow of Coupling constants}
\label{g1-g2}
}
\item
To check if  the universal curve is the renormalized trajectory which is
the continuum theory, let us study a scaling behavior.       
The continuum limit of the lattice theory is realized 
when $n\rightarrow\infty$ and $a\rightarrow 0$ for fixed $b$ is taken.
Effective actions are usually a function of the step $n$ of the
       block-spin transformation and the lattice distance
       $a(\beta,\kappa)$.
If the effective action shows a scaling, that is, a function of the
       product $b=na(\beta,\kappa)$ alone, it is the renormalized 
trajectory (RT). Since we have seen the couplings except the self one are
       well reproduced by the Coulomb interaction, let us concentrate on
       the constant coefficient of the Coulomb term in (\ref{lat-a_Coulomb}). 

The theory has two parameters $\beta$ and $\kappa$ on the lattice and 
$g$ and $v$ in the continuum. For the purpose of the comparison with the
       continuum theory, we tune $\beta$ and $\kappa$ so that we may get
       the same value of $v=\sqrt{\kappa/a(\beta,\kappa)}$ as in Table.\ref{v-beta}.
Here we fix $v\sim 2.56$. The values of the parameters $\beta$ and
       $\kappa$ 
in Table \ref{st_L} are chosen
       similarly.
\TABLE{
\begin{tabular}{|r||r|r||r|r||r|r|} \hline
  $\beta$~~ & {\bf $\kappa$}~~~~~ & {\bf $v$}~~~~~~~~
            & $\kappa$       ~~~~ & $v$       ~~~~~~~
            & $\kappa$       ~~~~ & $v$       ~~~~~~~ \\ \hline \hline
6.0 & 1.030 & {\bf 2.549(68)}  & &  & & \\ \hline
5.8 & 1.030 & {\bf 2.552(61)}  & 1.005 & 2.419(62) & 1.015 & 2.519(65)\\ \hline
5.6 & 1.035 & {\bf 2.554(66)}  & &  & & \\ \hline
5.4 & 1.055 & {\bf 2.566(62)}  & 1.050 & 2.443(67) & 1.065 & 2.656(65) \\ \hline
5.2 & 1.065 & {\bf 2.567(63)}  & 1.065 & 2.468(68) & 1.080 & 2.588(58) \\ \hline
5.0 & 1.065 & {\bf 2.549(66)}  & 1.095 & 2.621(63) & 1.100 & 2.704(60) \\ \hline
4.8 & 1.090 & {\bf 2.544(61)}  & 1.105 & 2.481(65) & 1.120 & 2.623(49) \\ \hline
4.6 & 1.140 & {\bf 2.549(65)}  & 1.150 & 2.602(64) & & \\ \hline
4.4 & 1.170 & {\bf 2.553(51)}  & 1.160 & 2.441(59) & 1.175 & 2.612(61) \\ \hline
4.2 & 1.210 & {\bf 2.548(50)}  & 1.195 & 2.515(62) & & \\ \hline
4.0 & 1.255 & {\bf 2.549(60)}  & 1.225 & 2.488(56) & & \\ \hline
\end{tabular}
\caption{Parameters of $\beta$ and $\kappa$ for fixed $v$.}
\label{v-beta}
}

Then we study in Fig.\ref{continuum-lim}
$n$ dependence for some $b$.
\FIGURE{
\epsfig{file=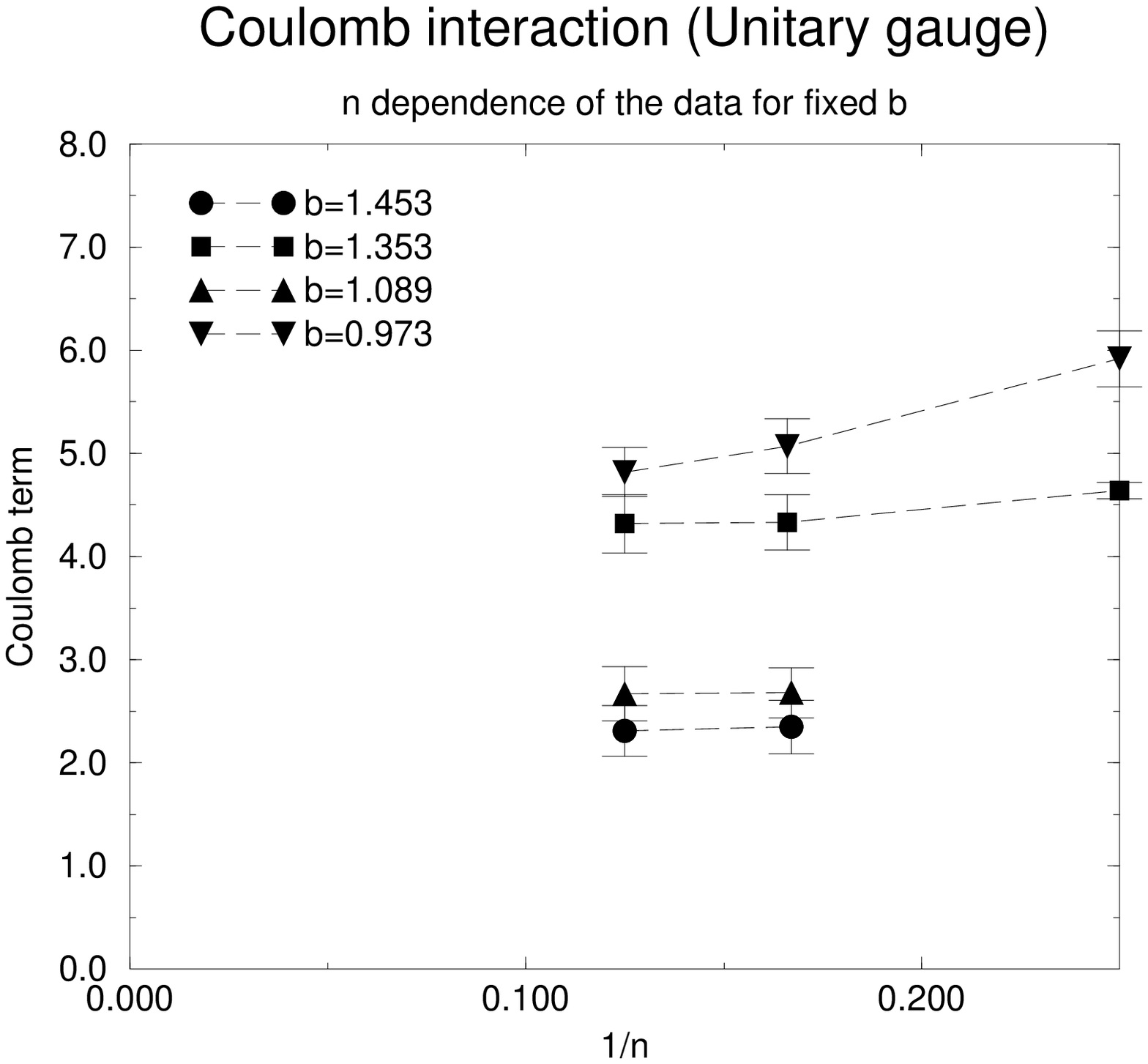,height=5cm}
\epsfig{file=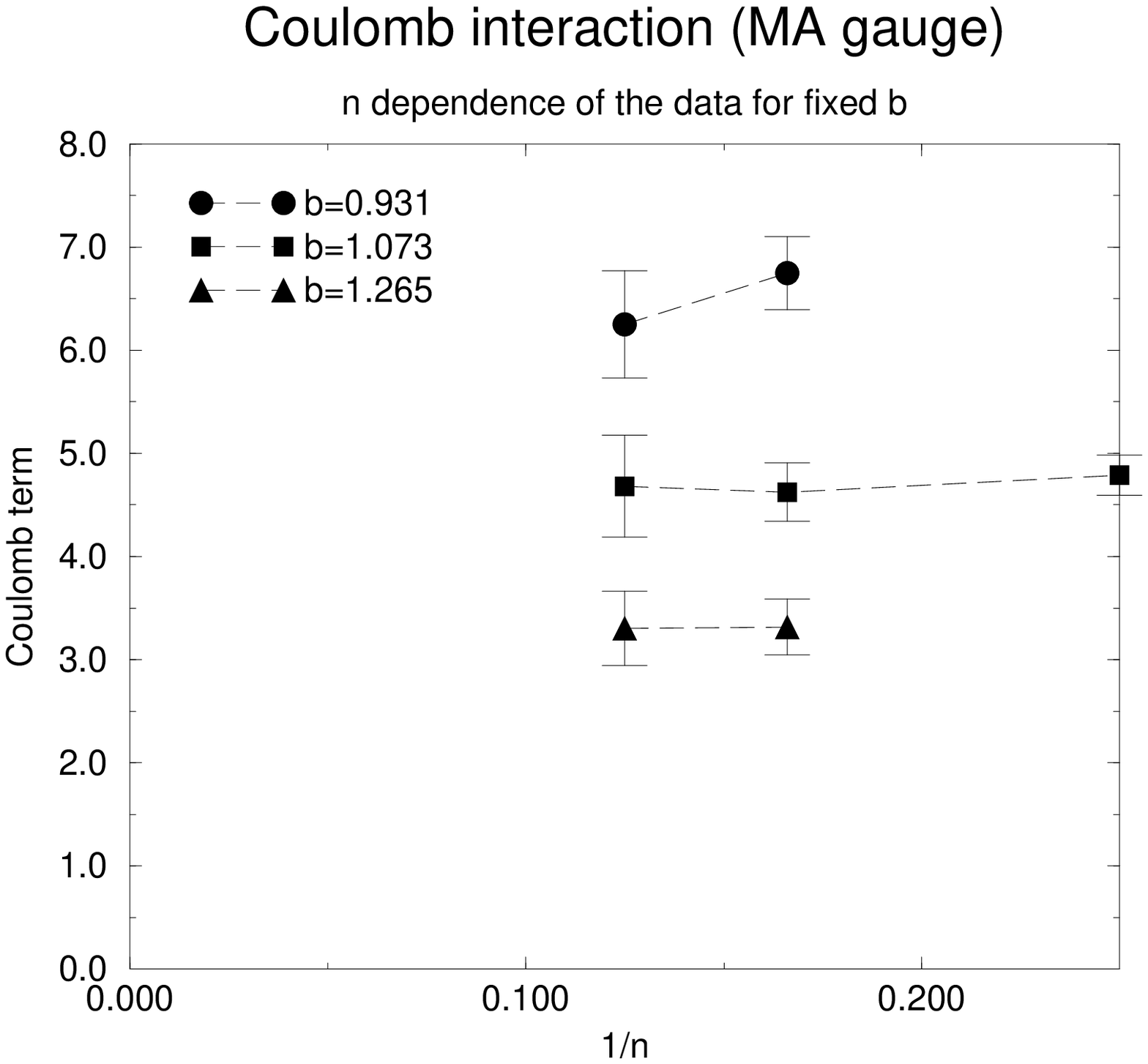,height=5cm}
\caption{$n$ dependence of Coulomb term}
\label{continuum-lim}
}
One can find that 
the action for $n\ge 4$ blocked instantons looks on RT. 
The scaling  can be seen more clearly in Fig.\ref{constf} in which we show 
the Coulomb term vs. $b$ for fixed $v\sim 2.56$.
This is very interesting.

\FIGURE{
\epsfig{file=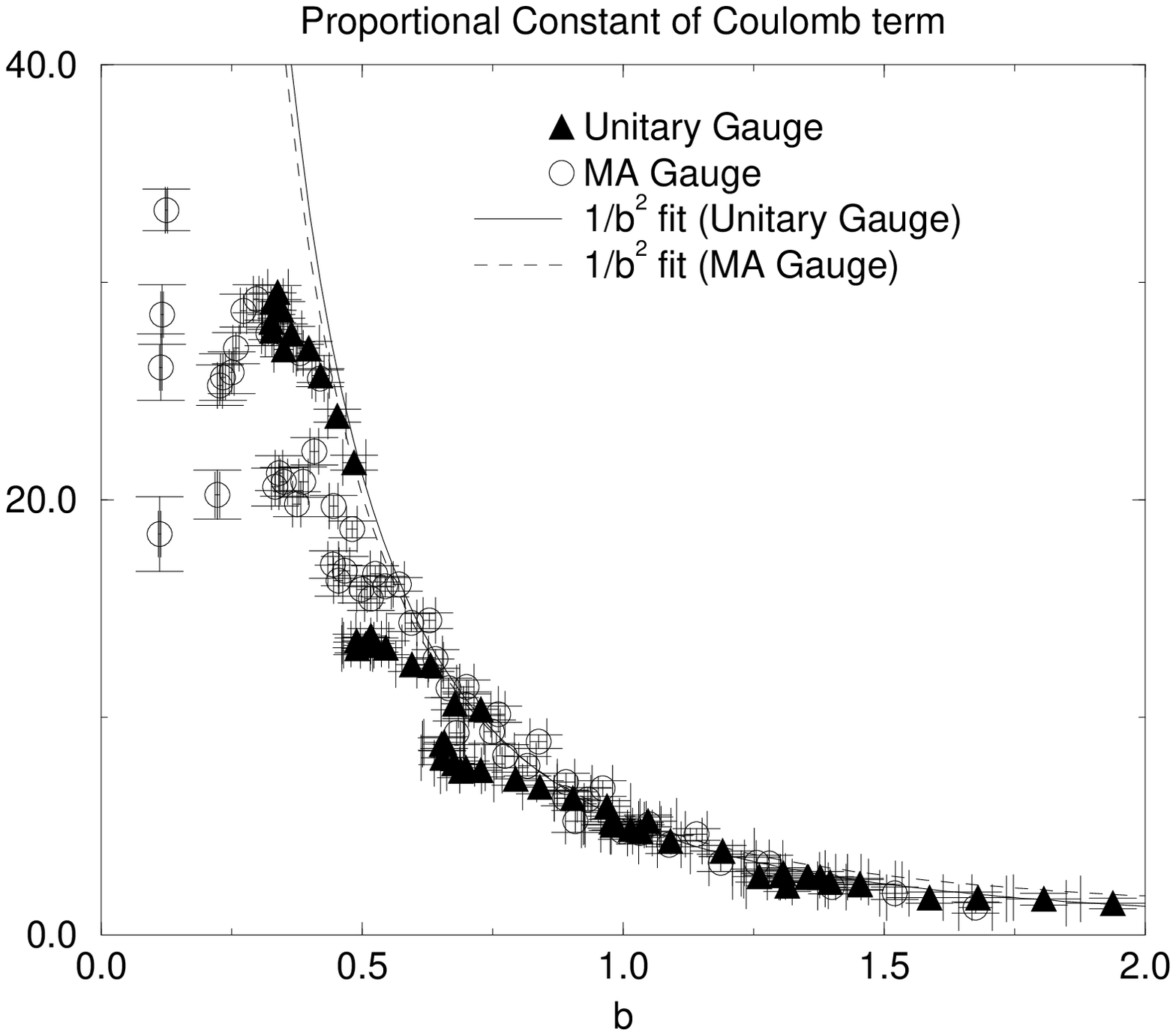,width=10cm}
\caption{Proportional coefficient of Coulomb term}
\label{constf}
}

\item
Let us determine the gauge coupling $g$ which is super-renormalizable in three
       dimensions. For that purpose, let us rewrite the action,
       restoring the lattice spacing $b=na$ explicitly:
\begin{eqnarray}
S[k]&=&\sum_{x,x'}b^6 k(bx) D(bx-bx') k(bx') \nonumber \\
    &=&{\rm Self ~ term}+
     \beta_m \sum_{x,x'}b^6 k(bx)\Delta_L^{-1}(bx-bx')k(bx') ,
\label{lat-instanton-action_b}
\end{eqnarray}
where $\beta_m$ is a constant. Since the mass dimension of the lattice
       Coulomb propagator is $1$ and that of $\delta^3(bx-bx')$ is $3$,
       the coefficients $C_i$ in (\ref{coeff-coulomb}) have the mass
       dimension $2$. $\beta_m$ is written as $8\pi/g^2$.
Hence the constant in (\ref{lat-a_Coulomb}) is expected to behave as 
\begin{equation}
{\rm Const.} \sim \frac{8\pi}{g^2}\frac{1}{b^2}.
\label{coef-const}
\end{equation}
The $1/b^2$ behavior in each gauge 
is actually observed for $b > 1$ as shown in 
Fig.\ref{constf}. 
>From the fitting (\ref{coef-const}), 
we can obtain the renormalized gauge coupling $g$ numerically.
In the case of $v\sim2.56$,
we find $g=3.9(2)$ in the unitary gauge condition and
$g=3.8(1)$ in MA gauge. Gauge independence is seen also in this case.
\end{enumerate}

\subsection{String tension of the Polyakov model}

Now let us calculate the string tension 
$\sigma_{\rm cl} =Mg^2/(4\pi)$
of Polyakov's model (\ref{st_an})
using the lattice parameters determined 
from the renormalized trajectory.

Note that
\begin{equation}
M^2=2\left(\frac{4\pi}{g}\right)^2
 N^{3/2} m_W^3 e^{-\frac{m_W}{g^2} \epsilon(\lambda/g^2) }
\end{equation} 
and 
\begin{equation}
N=\frac{m_W}{g^2}\alpha(\lambda/g^2).
\end{equation} 
As evaluated in Sec.\ref{GG}, we get 
$\epsilon(\infty)=4\pi\times 1.67$ and $\alpha(\infty)\simeq 4\pi \times 3.33$.
Also we have seen above that, 
for $v\sim2.56$,
$g=3.9(2)$ in the unitary gauge  and
$g=3.8(1)$ in MA gauge. 

Using these parameters, we can estimate the string tension from Eq.(\ref{st_an}).
In the case of the unitary gauge condition,
\begin{equation}
\sqrt{\frac{\sigma_{\rm cl}}{\sigma_L}} = 1.4(9)  ({\rm unitary ~ gauge}).
\end{equation}
Note that we are considering the scale using $\sigma_L$.
The error is not small,  since the errors 
of $g$ and $v$ is enlarged by the exponential factor 
in Eq.(\ref{zeta}).

In the case of MA gauge, we obtain
\begin{equation}
\sqrt{\frac{\sigma_{\rm cl}}{\sigma_L}} = 1.3(8)  ({\rm MA ~ gauge}).
\end{equation}

Considering the fact that the saddle point approximation used in
deriving Eq.(\ref{st_an}) is not perfectly justified for such small $g$
used here, the above agreement is rather surprising. In conclusion,
we may say that 
the confinement scenario of three-dimensional Georgi-Glashow model 
by Polyakov is thus verified  numerically.

\section{Concluding remarks}

We have studied three-dimensional lattice Georgi-Glashow model 
in the London limit 
$\lambda\to\infty$ 
under the restriction to an  intermediate region where the screening 
appears. In Ref.\cite{ambjorn}, the screening is seen around the range 
of $O(2m_W/\sigma)$. In our case this value is around 20 in our unit.
As seen in Fig.\ref{constf}, we are dealing with the region up to 
$b\sim 2$ in our unit. Hence the screening is not viable in this region. 

We have found that Polyakov's quasi-classical 
analyses on the basis of the dilute instanton gas approximation are 
in almost quantitative agreement with the Monte-Carlo results using 
the DeGrand-Toussaint definition of a lattice instanton both in the unitary 
and the MA gauges. 

There are other exciting topics to which one can apply the method
developed here.
\begin{enumerate}
 \item 
 When we integrate out non-zero modes of gauge and quark fields perturbatively in 
high temperature QCD in four dimensions, we get an effective model described in 
terms of the zero-mode bosonic fields alone. The model is equivalent to 
the three-dimensional Georgi-Glashow model \cite{Reisz}. 
The effective couplings $g$, $v$ and $\lambda$ are calculated in terms of 
the quantities in four-dimensional QCD. In this case we need to
perform simulations for the case with finite $\lambda$. 
It is very interesting to know if non-perturbative 
effects in the deconfinement phase of (QCD)$_4$ can be explained totally by 
the instanton effects. 
\item 
It is expected that, when $\lambda=0$, the three-dimensional 
Georgi-Glashow model
can not be described in terms of the Coulomb gas of instantons and the 
deconfinement phase will appear\cite{Magruder}. There is an attractive interaction
between instantons and anti-instantons which does not depend on the sign of 
the magnetic charge. It is a Yukawa-type interaction for $\lambda\ne 0$.
But when $\lambda=0$, the Higgs field becomes massless and the interaction just 
cancel the Coulomb repulsive interaction between the same-sign instantons or 
between anti-instantons. However, such a scenario needs information of the core 
region of instantons and so non-perturbative numerical evaluation is 
highly desirable.
\end{enumerate}

\acknowledgments

The authors thank Tomohiro Tsunemi for fruitful discussions.
This work is supported by the Supercomputer Project 
(No.98-33, No.99-47, No.00-59)
of High Energy Accelerator Research Organization (KEK) and
the Supercomputer Project of the Institute of Physical and 
Chemical Research (RIKEN).  T.S. acknowledges the financial support from
JSPS Grant-in Aid for Scientific Research (B) (Grant No. 10440073 and
No. 11695029)

\appendix

\section{Confinement from the instanton dilute gas model}
\label{string_tension}

String tension from the dilute gas model Eq.(\ref{st_an}) is
calculated as follows.

Let us start from  Eq.(\ref{partition_eta}):
\begin{eqnarray}
&&Z[\eta]=\int D\chi\exp{\left\{ -\frac12\left(\frac{g}{4\pi}\right)^2 \int{\rm d}^3x
  [(\partial_\mu(\chi-\eta) )^2 - 2M^2 \cos{\chi}] \right\} }, \\
&&\eta(x)=\int_S {\rm d}\vec{S}\cdot\frac{\vec{x}-\vec{y}}{|\vec{x}-\vec{y}|^3}
.
\end{eqnarray}

The path integral may be evaluated by the saddle point approximation 
when the coupling $g$ is large enough.
Then we have to 
solve the saddle point equation 
\begin{equation}
\partial^2(\chi_{\rm cl}-\eta)=M^2 \sin{\chi_{\rm cl}} .
\label{chi_eqmotion}
\end{equation}

We assume a Wilson loop that the contour $C$ is planar 
and is placed in the $xy$ plane
as in Fig.\ref{wilson-xyz} .
\FIGURE{
\epsfig{file=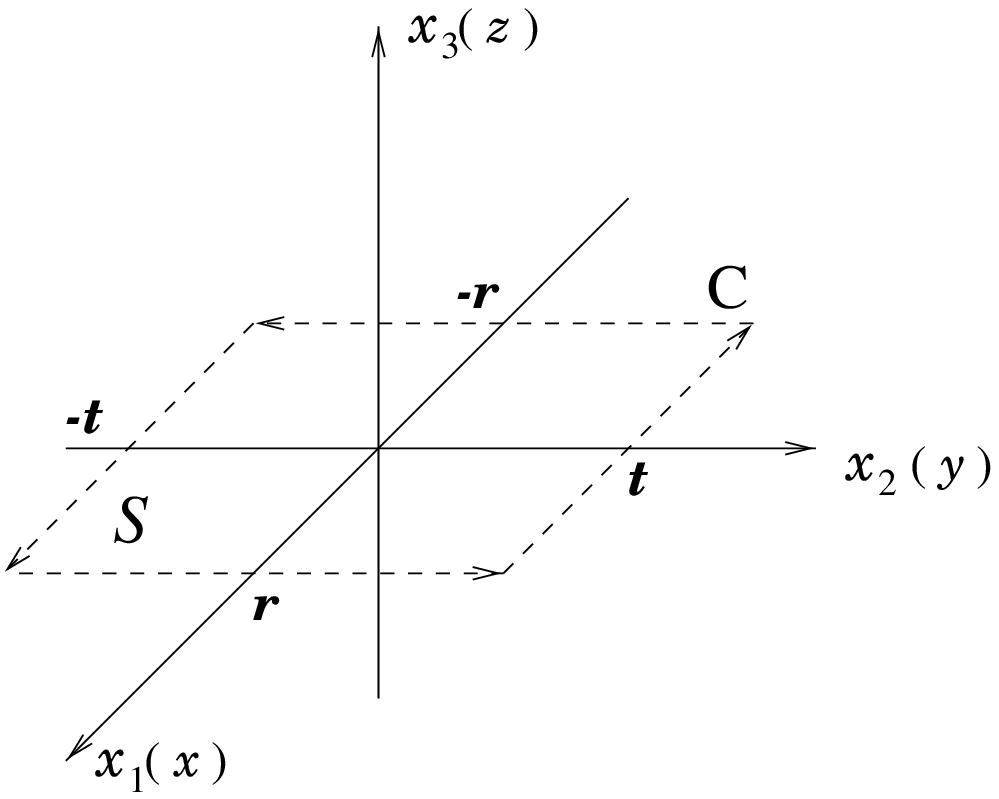,width=9cm,height=5cm}
\caption{Wilson loop}
\label{wilson-xyz}
}
Then Eq.(\ref{chi_eqmotion}) takes the form
\begin{equation}
\partial^2 \chi_{\rm cl}=2\pi\delta(z)\theta_S(xy)+M^2\sin{\chi_{\rm cl}}
,
\label{partition_eta_xyz}
\end{equation}
where
\begin{equation}
\theta_S(xy)=\left\{ \begin{array}{cl}
                      1 & ~~~~x, y \in S \\
                      0 & ~~~~{\rm otherwise}
                    \end{array} \right.
.
\label{theta_xy}
\end{equation}
Far from the boundaries of the contour,
Eq.(\ref{partition_eta_xyz}) is approximately one dimensional
($\chi_{\rm cl}$ depends only on $z$) and has a solution
\vspace{5mm}
\begin{equation}
\chi_{\rm cl}=\left\{ \begin{array}{cc}
         ~4\arctan(e^{-Mz}) & z > 0 \\
         -4\arctan(e^{Mz})  & z < 0 
                      \end{array} \right.
.
\label{chikai}
\end{equation}
$\eta(x)$ only depends on $z$ approximately, too.
Eq.(\ref{eta}) is rewritten as
\begin{eqnarray}
\eta(x) &=& \int_{-r}^r {\rm d}y_1 \int_{-t}^t {\rm d}y_2
   \frac{x_3}{\{ (x_1-y_1)^2 + (x_2-y_2)^2 + {x_3} ^2 \}^{3/2}} \nonumber \\
 &=& \arctan\left(
    \frac{(r-x_1)(t-x_2)}{x_3\sqrt{(x_1-r)^2+(x_2-t)^2+{x_3}^2}} \right)
 \nonumber \\
 &&   +\arctan\left(
    \frac{(r+x_1)(t-x_2)}{x_3\sqrt{(x_1+r)^2+(x_2-t)^2+{x_3}^2}} \right)
 \nonumber \\
 && 
    +\arctan\left(
    \frac{(r-x_1)(t+x_2)}{x_3\sqrt{(x_1-r)^2+(x_2+t)^2+{x_3}^2}} \right)
 \nonumber \\
 &&   +\arctan\left(
   \frac{(r+x_1)(t+x_2)}{x_3\sqrt{(x_1+r)^2+(x_2+t)^2+{x_3}^2}} \right) .
 \nonumber
\end{eqnarray}
Now, if $ r \gg x_1 $ and $t \gg x_2 $
( far from the boundaries of the contour),
it is reduced to
\begin{equation}
\eta(z) \sim 4\arctan\left(
    \frac{rt}{z\sqrt{r^2+t^2+z^2}} \right) .
\label{eta_z}
\end{equation}
Hence, one may estimate the integral of $Z[\eta]$ in terms of a
$z$ integral alone(where $x_3 \rightarrow z $ ):
\begin{equation}
Z[\eta] \sim \exp
 \left\{ \frac12 \left(\frac{g}{4\pi}\right)^2 S M^2 
   \int{\rm d}z 
  [(\chi_{\rm cl}(z)-\eta(z))\sin\chi_{\rm cl}+\cos\chi_{\rm cl}] \right\}
,
\label{z_eta_z}
\end{equation}
where $S$ is the area of the Wilson loop ($S=R\cdot T=4rt$).

>From Eq.(\ref{eta}), (\ref{z_eta_z}) and (\ref{eta_z}),
\begin{eqnarray}
W[C] &\sim& \exp\left\{-RT\left(\frac{g}{4\pi}\right)^2 M^2 \right.
  \nonumber \\
  && \left. \int_0^{\infty}{\rm d}z 
     \left[16\arctan\left(\frac{rt}{z\sqrt{r^2+t^2+z^2}} \right)
       \frac{e^{-Mz}(1-e^{-2Mz})}{(1+e^{-2Mz})^2} \right] \right\}
.
\label{wilson_math}
\end{eqnarray}
The string tension is given by
\begin{equation}
\sigma_{\rm cl}=M\left(\frac{g}{\pi}\right)^2
  \int_0^{\infty}{\rm d}z 
  \left[\arctan\left(\frac{M^2rt}{z\sqrt{(Mr)^2+(Mt)^2+z^2}} \right)
    \frac{e^{-z}(1-e^{-2z})}{(1+e^{-2z})^2} \right]  .
\label{st_math}
\end{equation}
The $z$ integral of Eq.(\ref{st_math}) is finite 
if $Mr,Mz \rightarrow \infty$ 
.
Thus we obtain the string tension Eq.(\ref{st_an})
\begin{equation}
\sigma_{\rm cl}=\frac{M g^2}{4\pi}.
\end{equation}

\section{Swendsen's method}
\label{swend}

We determine a lattice instanton action from
instanton configurations on three-dimensional lattice.
A theory of instanton is given in general by the following partition
function
\begin{equation}
Z=\left(\prod_{x} \sum_{k(x)=-\infty}^{+\infty} \right)~e^{-S[k]},
\end{equation}
where 
$S[k]$ is an instanton action describing the theory.
Consider a set of all independent operators which are
summed up over the whole lattice.
We denote each operator as $S_i[k]$.
Then, the action may be written as a linear combination of these operators,
\begin{equation}
S[k]=\sum_i G_i S_i[k],
\label{instant-action1}
\end{equation}
where
$G_i$ are renormalized coupling constants.

The expectation value of operator ${\cal O}[k]$ is estimated as
\begin{equation}
<{\cal O}[k]> = \frac{1}{Z}
 \left(\prod_{x} \sum_{k(x)=-\infty}^{+\infty} \right)
 {\cal O}[k] ~e^{-S[k]} .
\label{exp-Z}
\end{equation}
Here, we focus on a certain site $x'$.
We define $\hat{S}[k]$ which contains $k(x')$ and rewrite
the numerator of Eq.(\ref{exp-Z}) as
\begin{eqnarray}
&&\left(\prod_{x}\sum_{k(x)=-\infty}^{+\infty} \right){\cal O}[k] e^{-S[k]}
 \\
&&=\left({\prod_{x}}'\sum_{k(x)=-\infty}^{+\infty} \right)
 ~ e^{-\sum_i G_i (S_i[k]-\hat{S}_i[k])}
 \{ \sum_{k(x')=-\infty}^{+\infty} {\cal O}[k] ~e^{-\sum_i G_i \hat{S}_i[k]} \} 
,
\label{exp-Z21}
\end{eqnarray}
where ${\prod}'$ means the product except for the site $x'$.
We rewrite Eq.(\ref{exp-Z21}) as
\begin{eqnarray}
(\ref{exp-Z21})&=&\left({\prod_{x}}'\sum_{k(x)=-\infty}^{+\infty} \right)
 ~ e^{-\sum_i G_i (S_i[k]-\hat{S}_i[k])}
 \nonumber \\
 && ~~~\sum_{k(x')=-\infty}^{+\infty}
 \frac{\sum_{\hat{k}=-\infty}^{+\infty} {\cal O}[\hat{k},\{k\}']~
  e^{-\sum_i G_i \hat{S}_i[\hat{k},\{k\}']} }
  {\sum_{\hat{k}=-\infty}^{+\infty}e^{-\sum_i G_i \hat{S}_i[\hat{k},\{k\}']} }
 ~~e^{-\sum_i G_i \hat{S}_i[k] } \nonumber \\
 &=&\left(\prod_{x}\sum_{k(x)=-\infty}^{+\infty} \right)
 \hat{\cal O}[k]~ e^{-\sum_i G_i S_i[k]}
,
\label{exp-Z3}
\end{eqnarray}
where 
\begin{equation}
\hat{\cal O}[k]=
\frac{\sum_{\hat{k}=-\infty}^{+\infty} {\cal O}[\hat{k},\{k\}']~
  e^{-\sum_i G_i \hat{S}_i[\hat{k},\{k\}']} }
  {\sum_{\hat{k}=-\infty}^{+\infty}e^{-\sum_i G_i \hat{S}_i[\hat{k},\{k\}']} }
,
\label{exp-Z4}
\end{equation}
$\hat{k}=k(x')$ and $\{k\}'$ is the coset of $\hat{k}$.

When we consider the DeGrand-Toussaint definition of instanton,
the sum with respect to instanton charge in Eq.(\ref{exp-Z4})
is not $[-\infty , +\infty]$ but to $[-(3n^2-1), 3n^2-1]$, where
$n$ is a step of  block spin transformation of  instanton.

Hence we get an identity 
\begin{equation}
<{\cal O}[k]>=<\hat{\cal O}[k]>.
\end{equation}

Since we don't know the correct set of coupling constants $G_i$, we have to
start from a trial set of coupling constants $\tilde{G}_i$ . 
We define $\overline{\cal O}[k] $ in which 
the true coupling constants $G_i$ in Eq.(\ref{exp-Z4})
are replaced by the trial ones $\tilde{G}_i$ as
\begin{equation}
\overline{\cal O}[k]=
\frac{\sum_{\hat{k}=k_{\rm min}}^{k_{\rm max}} {\cal O}[\hat{k},\{k\}']
  ~ e^{-\sum_i \tilde{G}_i \hat{S}_i[\hat{k},\{k\}']} }
 {\sum_{\hat{k}=k_{\rm min}}^{k_{\rm max}}
   ~ e^{-\sum_i \tilde{G}_i \hat{S}_i[\hat{k},\{k\}']} }
.
\label{bar_O}
\end{equation}
If the trial $\tilde{G}_i$ are not equal to the true $G_i$,
$<{\cal O}[k] - \overline{\cal O}[k] > \neq 0 $.
When $\{ \tilde{G}_i \}$ are not far from $\{ G_i \}$,
we get 
\begin{equation}
<{\cal O}[k] - \overline{\cal O}[k] >
\sim\sum_i < \overline{\cal O}~\overline{S}_i - (\overline{{\cal O}S_i})>
  (G_i-\tilde{G}_i) .
\label{gi-gi}
\end{equation}
Actually, if we use $S_i[k]$ as an operator ${\cal O}[k]$,
we get a 
very good convergence .
Hence we use here
\begin{equation}
<S_i[k] - \overline{S}_i[k] >
 \sim \sum_j < \overline{S}_i~ \overline{S}_j - (\overline{S_i S_j})>
  (G_j-\tilde{G}_j) .
\label{swd-eq}
\end{equation}
Since the value of $S_i[k]$ is obtained from vacuum configurations,
we can solve Eq.(\ref{swd-eq}) iteratively to obtain the real 
set of coupling constants $G_i$.


\end{document}